\documentclass[review]{elsarticle}
\pdfoutput=1

\usepackage{lineno,hyperref}
\modulolinenumbers[5]

\journal{}

\usepackage{float}
\usepackage{subcaption}
\usepackage{mathtools}
\usepackage[]{algorithm2e}
\usepackage{amssymb}
\usepackage[framemethod=tikz]{mdframed}
\usepackage{setspace} 
\usepackage{nomencl}
\usepackage{glossaries}
\usepackage{moreverb}
\usepackage{lineno}
\usepackage{comment}
\usepackage{booktabs}

\usepackage{xcolor}
\usepackage{framed}
\usepackage{lipsum}
\usepackage[margin=0.7in]{geometry}

\colorlet{shadecolor}{blue!20}
\usepackage{nomencl}
\makenomenclature

\begin{document}

\begin{frontmatter}

\title{Integrating Discrete Sub-grid Filters with Discretization-Corrected Particle Strength Exchange Method for High Reynolds Number Flow Simulations}




\author[mymainaddress]{Anas Obeidat\corref{mycorrespondingauthr}}
\cortext[mycorrespondingauthr]{Corresponding author}
\ead{anas.obeidat@uni.lu}


\address[mymainaddress]{Department of Engineering, Institute of Computational Engineering, University of Luxembourg, Luxembourg}

\begin{abstract}
We present a discrete filter for subgrid-scale (SGS) model, coupled with the discretization corrected particle strength exchange (DC-PSE) method for the simulation of three-dimensional viscous incompressible flow, at high Reynolds flows. 
The majority of turbulence modelling techniques, particularly in complex geometries, face significant computational challenges due to the difficulties in implementing 3-D convolution operations for asymmetric boundary conditions or curved domain boundaries.

In this contribution Taylor expansion is used to define differential operators corresponding to the convolution filter, so that the transfer function remains very close to the unity of sizeable displacement in wave number, making the filter a good approximation to the convolution one. A discrete Gaussian filter, in both fourth and second-order forms, was evaluated with varying ratios of particle spacing to the cut-off length. The impact of the filter's order and the ratio's value is thoroughly examined and detailed in the study.
Additionally, the Brinkman penalisation technique is employed to impose boundary conditions implicitly, allowing for efficient and accurate flow simulations around complex geometries without the need for modifying the numerical method or computational domain. 
The incompressible flow is governed by the the he Entropically Damped Artificial Compressibility equations allowing explicit simulation of the incompressible Navier–Stokes equations.

The effectiveness of the proposed methodology is validated through several benchmark problems, including isotropic turbulence decay, and flow around four cylinders arranged in a square in-line configuration. These test cases demonstrate the method's accuracy in capturing the intricate flow structures characteristic of high Reynolds number flows, highlighting its applicability to industrial turbulence modeling.

\end{abstract}


\end{frontmatter}

\section{Introduction}
Large eddy simulation (LES)~\cite{Rogallo:1984, Sagaut:1998} is a widely used technique for modeling intricate, transient flow dynamics. The LES governing equations are derived by modifying the Navier-Stokes equations through the application of a low-pass spatial filter, thereby computing solutions for the filtered quantities. Such a process effectively separates the flow into resolved (low-frequency) and unresolved (high-frequency) components. The latter, known as subgrid-scale (SGS) modes, are characterised using the subgrid-scale model.
Numerous subgrid-scale (SGS) models have been introduced, it is agreed that that the models which analyze the highest frequency resolved modes demonstrate the greatest efficiency~\cite{Sagaut:1999}. These modes are derived from the resolved flow through the application of a low-pass filter, known as a test filter.

The most celebrated model is the Smagorinsky model~\cite{Smagorinsky:1963}, assuming that the eddy viscosity (added viscosity until the viscous scales are fully resolved through the discretization process) is determined based on the magnitude of the resolved strain rate alongside characteristic length scales. These scales are proportional to the filter width,adjusted by a specific Smagorinsky constant. Research has shown that Smagorinsky constant is not a single value, but rather fluctuates based on the computational resolution of the flow's characteristics, with variations ranging from $0.1$ to $1.0$~\cite{Smagorinsky:1993,Pope:2012}. Germano et al. and Lilly~\cite{Germano:1991,Lilly:1992} introduced the Germano Lilly dynamic procedure, which allows the Smagorinsky constant to change over time. Ghosal et al.~\cite{Ghosal:1995} introduced the local dynamic procedure, enabling the Smagorinsky constant to vary both temporally and spatially.

Bardina et al.\cite{Bardina:1980} established a connection between the model's kinetic energy amplitude and the test field, leading to the development of the turbulent kinetic energy (TKE) model. Sagaut and Loc~\cite{Sagaut:1996} extended this concept through presenting the mixed scale model (MSM),  which combine principles from both  the TKE model and Smagorinsky model~\cite{Smagorinsky:1963}. Lesieur and  M\'etais proposed an improvement to their structure function model~\cite{Metais:1992} (the selective structure function mode) by incorporating a test field to examine the vorticity's topology~\cite{Lesieur:1996}. Liu et al.~\cite{Liu:1994} expanded this methodology by applying the same LES filter and test filter to the case of different cutoff wave numbers, distinguishing between two levels of filtering (a finer and a coarser one).

In the context of turbulent flow simulations in complex/industrial geometries, challenges arise with the selection of an suitable test filter. The SGS models were established theoretically by defining the test filter operation as convolution products that combine the velocity field with the filter kernel. This definition aligns well with numerical approaches like the pseudo-spectral method~\cite{Sagaut:1999}. However, for local approximation methods, which include finite differences, finite volumes, and finite elements—methods widely utilised in industrial applications—the implementation of $3$-D convolution becomes complex and computationally expensive (e.g. for application of asymmetric boundary conditions, curved boundaries of the domain).

Due to the aforementioned challenges, a discrete test filter employing compact stencils based on weighted averages is often used.
Vasilyev et al.~\cite{Vasilyev:1998} presented a set of rules for constructing discrete filters in complex geometries, with the aim to minimise the projection error between the numerical differentiation and filtering processes. Vasilyev demonstrated that  discrete filters do not inherit the properties of the conventional continuous filters. Therefore, to maintain consistency between the discrete form and the SGS model, selecting a suitable discrete filter along with its specific properties becomes essential.

Mesh-free methods represent continuous fields using a collection of points (such as particles or point clouds) situated at specific locations, without the need for connectivity between them. Among these methods, the Smoothed Particle Hydrodynamics (SPH) method is widely used approach for modeling turbulence. First developed by Gingold and Monaghan~\cite{Gingold:1977} and Lucy~\cite{Lucy:1977}. Monaghan~\cite{Monaghan:2002} adapted SPH for two-dimensional turbulence modeling. Violeau and Issa~\cite{Violeau:2007} introduced three SPH turbulence models, and Dalrymple and Rogers~\cite{Dalrymple:2006} applied an LES model for wave simulations with SPH. Robinson and Monaghan~\cite{Robinson:2012} examined SPH's efficacy in DNS of decaying turbulence. Challenges in SPH turbulence modeling include lack of physical viscosity and particle clustering, leading to potential inaccuracies. Borve~\cite{Borve:2001, Borve:2005} improved SPH's precision with a regularised approach, and Chaniotis et al.\cite{Chaniotis:2002, Chaniotis:2003} developed remeshed SPH (rSPH) for uniform particle distribution. Obeidat and Bordas~\cite{Obeidat:2017} proposed a hybrid rSPH method, combining Lagrangian and Eulerian advantages for enhanced modeling accuracy and then added a penalised term for flow simulation complex/industrial geometries~\cite{Obeidat:2019,Obeidat:2020}.

The Discretization-Corrected particle strength exchange (DC-PSE) method, which generalises finite difference methods to meshfree point clouds, was first introduced by Schrader et al.~\cite{Schrader:2010}. The aim of DC-PSE method is to eliminate the discretization error by deriving the kernels directly from the set of discrete moment conditions evaluated on the set of point clouds, when satisfied, guarantee the approximation of a specified differential operator with a determined order of convergence.

The Entropically Damped Artificial Compressibility (EDAC) formulation was introduced by ~\cite{Clausen:2013}. The EDAC formulation allows explicit simulation of the incompressible Navier–Stokes (INS) equations, by damping pressure waves via an entropy generation mechanism. The explicit formulation of the INS avoids the computationally expensive solution of a global Poisson equation and simplify the implementation.
The EDAC formulation was effectively applied to model viscous flow at high Reynolds number References~\cite{ClausenB:2013,Vermeire:2024,Kajzer:2018,Trojak:2022}.
For our knowledge the DC-PSE method with EDAC formulation has not been implemented to model viscous flow at high Reynolds numbers nor to model turbulence flow with large eddy simulation model.

In this paper we present a integrated DC-PSE method with discrete filter. The computational approach introduced in~\cite{Singh:2023} is extended to the simulation of viscous compressible high Reynolds flows. To accurately capture the turbulent structures characteristic of high Reynolds number flows, the particle-based numerical approach is adapted to large eddy simulation for turbulence modeling using a discrete subgrid scale (SGS) filters with different orders.

\section{Discrete approximation of the continuous filter }
\label{sec2}
Large eddy simulation (LES) is based on the principle of separating the flow into its low-frequency (large-scale) and high-frequency (small-scale) components. The low frequencies frequency components are directly resolved through discretization, whereas the high frequency components known as subgrid-scale (SGS) modes are modeled implicitly. 

Ideally, the large scales should represent the part of the spectrum where mostly energy transfer takes place, while the small scales are primarily associated with dissipative processes. 
The effectiveness of LES in accurately capturing these dynamics  depends on the actual grid resolution, which directly influences the quality of scale separation and, consequently, the success of the simulation method.

For example, the field $f$ is decomposed into the resolved and filtered components, as follows
\begin{equation}
f(x,t)=\bar f(x,t)+f'(x,t),
\end{equation}
where $\bar f$ is the resolved (filtered or larger) component, $f'$ donates the subgrid component, $x$ is the spatial coordinates and $t$ is time.
The resolved component $\bar f$ within the computation domain $\Omega$ is defined through a convolution product between a frequency low-pass filter kernel $G(x,\xi,\Delta)$ and the flow variable $f$ with a spatial filter width $\Delta$ as:

\begin{equation}
\label{convo_eq}
\bar{f}(x,t) =\int_{\Omega}^{}G(x,\xi,\Delta) f(\xi,t) d\xi ,
\end{equation}
where $x$ represents the coordinates of the point where the filter is applied, $\xi$ is the variable of integration over the spatial domain.
This equation establishes the mathematical relationship for filtering the flow variable $f$ over the computational domain $\Omega$, using a low-pass filter kernel $G$ with a spatial filter width $\Delta$, to separate the larger, resolved scales from the smaller, subgrid scales.

\subsection{Discrete  numerical filters}
One of the most used filters in LES is the Gaussian filter, defined as follows:
\begin{equation}
G(x-y) =\sqrt\frac{6}{\pi\bar\Delta^2}exp(-\frac{6 \left | x-y  \right |^2}{\bar\Delta^2}),
\end{equation}

where $G(x-y)$ represents the filter kernel between points $x$ and $y$, and $\bar\Delta$ is the characteristic filter width.

Taylor expansion is applied to define the differential operators associated with the convolution filter, ensuring  that the transfer function closely approximates unity for sizeable displacements in wave number. This approach makes the filter an effective approximation of the convolution one. Sagaut and Grohens~\cite{Sagaut:1999} derived the discrete operators for the Gaussian filter as:

Sagaut and Grohens~\cite{Sagaut:1999}, derived the discrete operators specifically for the Gaussian filter and box filter as follows:

\begin{itemize}
\item Gaussian filter of second order ($\textbf{G-O(2)}$),
\\ 
\begin{equation}
\label{gauss2}
\bar f_i = \frac{1}{24} \epsilon^2 (f_{i+1} +f_{i-1})+\frac{1}{12}(12-\epsilon^2)f_i ,
\end{equation}

\item Gaussian filter of fourth order ($\textbf{G-O(4)}$),
\\
\begin{equation}
\bar f_i =\frac{\epsilon^4- 4\epsilon^2}{1152}(f_{i+2}+f_{i-2})+\frac{16\epsilon^2- 4\epsilon^4}{228}(f_{i+1}+f_{i-1})+
\frac{\epsilon^4- 20\epsilon^2+192}{129}f_i ,
\label{epsilon}
\end{equation}

%
\end{itemize}
where $\epsilon$ is the ratio of the mesh size to the cut-off length scales of the targeted filter, corresponding to the smallest resolved wave number  $k_c$. The presented discretisation ensures that the deferential operator is equivalent to the target filter up to the $n$th order.

The filters described can be adapted for multi-dimensional applications by employing either a linear combination or a product-based construction, expressed as $F^{dim}=\frac{1}{dim}\sum_{i=1}^{dim}F^i$ for linear combinations, and $F^{dim}=\frac{1}{dim}\prod _{i=1}^{dim}F^i$ for product constructions. Here, $F^i$ is the one-dimensional filter defined applied along the $i$th spatial direction, and $dim$ is the spatial dimensions. 

Now, for filtering a quantity $f$ using a one-dimensional filter $F$ with $N$ point stencil support we get,
\begin{equation}
\bar f_{i,j,k}=  \sum_{l=-N}^{N} a_l (f_{i+l,j,k}+f_{i,j+l,k}+f_{i,j,k+l})
\end{equation}

For the $3$-dimensional case with an $N$-point stencil, the filtered quantity is obtained through:
\begin{equation}
\bar f_{i,j,k}=   \sum_{l=-N}^{N}  \sum_{n=-N}^{N}  \sum_{m=-N}^{N} a_l a_n a_m (f_{i+l,j,k} f_{i,j+n,k}+f_{i,j,k+m}).
\end{equation}

In this study, both second and fourth order filters are applied to one of the test cases in section\ref{turbo}, providing a concise analysis of the filter order's impact on the solution.  For further analysis please refer to~\cite{Sagaut:1999}.
Later the fourth order is specifically applied to the study involving flow around a cylinder~\ref{flow-cylinder}.

\section{The Filtered Entropically Damped Artificial Compressibility Equations}
\vspace{-2pt}
\label{GE}

The EDAC formulation introduces a pressure evolution equation, derived form the thermodynamics of the system under a constant density.
The dynamics of the flow  is governed by the filtered EDAC equations, which outline the conservation of momentum and the evolution of pressure as follows:

\begin{eqnarray}
\rho\frac{d\bar{u}_i}{dt} + u_j \frac{\partial \bar{u}_i}{\partial x_j}&=&   - \frac{\partial p}{\partial x_i}  +  
                         \frac{\partial \tau_{ij}}{\partial x_j} +
                          \frac{\partial \tau^\mathit{sgs}_{ij}}{\partial x_j}
\label{momentum-euler}
\\
\frac{d p}{dt}  +  \bar{u}_i\frac{\partial p}{\partial x_i} &=&  -{c_s}^2 \rho_o   \frac{\partial \bar{u}_i}{\partial x_i} + \nu \frac{\partial^2 p}{\partial x_i x_i}. 
\end{eqnarray}
The shear stress $\tau_{ij}$ is given by:
\begin{equation}
   \label{shearstress}
   \tau_{ij} = \mu 
      \left( \frac{\partial u_i}{\partial x_j} + 
             \frac{\partial u_j}{\partial x_i} - 
             \frac{2}{3} \delta_{ij} \frac{\partial u_k}{\partial x_k} 
      \right),
\end{equation}
and the sub-grid scale stress $\tau^\mathit{sgs}_{ij}$ is defined as:
\begin{equation}
   \tau^\mathit{sgs}_{ij}=  \overline {u_iu_j} -\bar{u}_i \bar{u}_j
   \label{sgs1}
\end{equation}

where $(\overline{\diamond})$ is the filtered variable, which can be expressed as a convolution product by applying Eq.~\ref{convo_eq}, or by applying the discreet filter as in Eqs.(~\ref{gauss2},~\ref{epsilon}), the reset of the quantities are defined as, $\bar{u}_i$ is the filtered velocity, $p$ is the  pressure, $\mu$ is the dynamic viscosity, $c_{s}$ is the speed of sound, $\tau_{ij}$ is the shear stress defined in Eq.\ref{shearstress}, $\delta_{ij}$ is the Kronecker delta, and $\tau^\mathit{sgs}_{ij}$ is the sub-grid stress tensor and defined in Eq.~\ref{sgs1}

The SGS interacts between the resolved (grid) scales and the unresolved (subgrid) scales. In this study, we employ the Smagorinsky model~\cite{Smagorinsky:1963}, to model the turbulent sub-grid stresses. This model, an eddy viscosity model is defined as:
\begin{equation}
   \tau^\mathit{sgs}_{ij}= \nu_{sgs}\bar{S}_{ij},
   \label{sgs}
\end{equation}
which links the sub-grid stress $\tau_{sgs}$ to the eddy viscosity $\nu_{sgs}$ and the resolved-scale strain rate tensor ${S}_{ij}$.
The eddy viscosity is defined as by:
\begin{equation}
\nu_{sgs} =\rho \left ( C_{s}  \Delta  \right )^{2} \sqrt{2\bar{S}_{ij}\bar{S}_{ij}},
\end{equation}
where $C_{s}$ is a non dimensional constant, with values in the literature~\cite{Rogallo:1984} suggested to range from $0.1$ to $1.0$, and  $\Delta$ is the model length scale $=\left ( \Delta_x  \Delta_y  \Delta_z \right )^{1/3}$.

The resolved-scale strain rate tensor ${S}_{ij}$ is calculated by:
\begin{equation}
 \bar{{S}}_{ij} = \frac{1}{2}\left(\frac{\partial \bar{u}_{i}}{\partial x_{j}} + 
                                                    \frac{\partial \bar{u}_{j}}{\partial x_{i} }\right),
\end{equation}
providing a framework for capturing the dynamics of turbulent flows within the sub-grid scale by relating the stresses to the strain rates experienced by the fluid at the resolved scale.

In the presented study, the flow is uniquely characterised by the Reynolds number 
$\mathit{Re}=U_o\rho_0 L /\mu$, and  the Mach number $\mathit{Ma} = U_o/cs$. $L$ is the characteristic length, $\rho_0$ is the reference density, $U_o$ is the reference velocity.

For applications requiring numerical simulations of viscous flows in complex/industrial geometries, the governing equations are coupled with Brinkman penalisation technique, as presented in~\cite{Obeidat:2019}. The computational domain is implicitly penalised by a mask function$\chi$ marking the regions where the solid geometry $O$ is located as:
\begin{equation}
 \label{chi-eq}
\chi(x) =\left\{\begin{matrix}
1 & \text{if} \, \, \, x \in O,\\ 
 0&  \text{otherwise}.& 
\end{matrix}\right.
 \end{equation}
 
To account for the presence of solid obstacles within the flow, a penalty term is incorporated into the momentum equations, modifying them as follows:
\begin{eqnarray}
\rho\frac{d\bar{u}_i}{dt} + u_j \frac{\partial \bar{u}_i}{\partial x_j}&=&   - \frac{\partial p}{\partial x_i}  +  
                         \frac{\partial \tau_{ij}}{\partial x_j} +
                          \frac{\partial \tau^\mathit{sgs}_{ij}}{\partial x_j}
                            - \frac{\chi}{\eta}( \bar{u}_i - u_{(oq)i}),
\label{momentum-pen}
\end{eqnarray}
where $\bar{u}_{(oq)i}$ represents the velocity of the solid body, $\phi$ is the porosity and $\eta = \alpha \phi$ is the normalised viscous permeability. It is important to note that both 
$\phi$ and $\eta$ are very small positive quantities, indicating low porosity and permeability, respectively. The mask function $\chi$, serving as a Heaviside function, distinguishes between the fluid and solid regions, thereby effectively applying the penalization where needed. 


\section{ The Discretization Corrected Particle Strength Exchange method (DC-PSE)}
The Discretization-Corrected Particle Strength Exchange (DC-PSE) method is a computational approach designed for discretizing differential operators on regular or irregular point clouds~\cite{Schrader:2010}. DC-PSE method employs a symmetric smoothing kernel $\eta \epsilon()$, to accurately approximate continuous functions $f_\epsilon(\vec{x})$ that are sufficiently smooth. The approximation is mathematically formulated as follows:
\begin{align}
		f(\vec{x}) \approx f_{\epsilon}(\vec{x})=\int_{\Omega} f(\vec{y}) \eta_{\epsilon}(\vec{x}-\vec{y}) \mathrm{d} \vec{y},
	\end{align}
 where $\epsilon$ is the smoothing length (kernel width).
 In DC-PSE, a linear system is solved for each particle locally to compute kernel weights, ensuring they meet the discrete moment conditions for a targeted order of convergence. This process allows DC-PSE to adhere to these conditions directly, based on particle distribution, thus significantly reducing quadrature errors.
We invite the reader to explore~\cite{Singh:2021} for a more in-depth examination of the EDAC formulation in conjunction with the DC-PSE method. This study offers validation and an extensive analysis of both convergence and accuracy related to the method 
\section{Numerical verification}
To verify the effect of the discrete filter, we deployed various numerical tests focusing on its performance across different computational benchmarks. These tests were designed to assess how the discrete filter influences the resolution of turbulence, especially in simulations with high Reynolds numbers.
For all the test cases the flow is characterized by the non-dimensional Mach number $\mathit{Ma}$, the Reynolds number $\mathit{Re}$, and the flow quantities $u$, $p$ and $\rho$ are normalized by either the maximum or the reference corresponding quantity. 
\subsection{Three-dimensional Taylor green vortex turbulence decay}
\label{3DTG}

\subsection{Three-dimensional isotropic turbulence decay}
\label{turbo}
Initial conditions for three-dimensional isotropic turbulence are obtained from the Johns Hopkins Turbulence Database (JHTDB)~\cite{Li:2008b}, based on direct numerical simulation data of forced isotropic turbulence across a $1024^3$ grid. The statistical characteristics of turbulence include a dissipation rate $\varepsilon =0.103$ derived from  $\varepsilon=2 \nu\left \langle S_{ij}S_{ij} \right \rangle$, an rms velocity $u'= \sqrt{(2/3)E_{tot}}=0.8$, with $E_{tot}=1/2 \left \langle u_i u_i\right \rangle $ representing the total kinetic energy and a Taylor-scale Reynolds number $\mathit{Re}_\lambda= u'\lambda/\nu=500$, calculated with $\lambda =\sqrt{15 \nu u^{'2}/\varepsilon} =0.13$ as the Taylor micro scale.

The database provide three velocity components and the pressure for incompressible flow, with data detailed for $1024$ elements in each spatial direction.

Date resolution at $1024^3$ is preprocessed first using a Gaussian filter with a specified cutoff to reduce the noise, and subsequently downsampled to the desired resolution -number of point clouds in each coordinate direction-here, $64^3$ $128^3$ and $256^3$. Through this preprocessing, a set of initial conditions is established for the benchmarking, consisting of pressure and velocity fields alongside particle coordinates in a $3$-dimensional domain scaled to $64, 128$ and $256$ point cloud in each coordinate direction.

\begin{figure}[H]
\center
     \includegraphics[clip, width=1\textwidth]{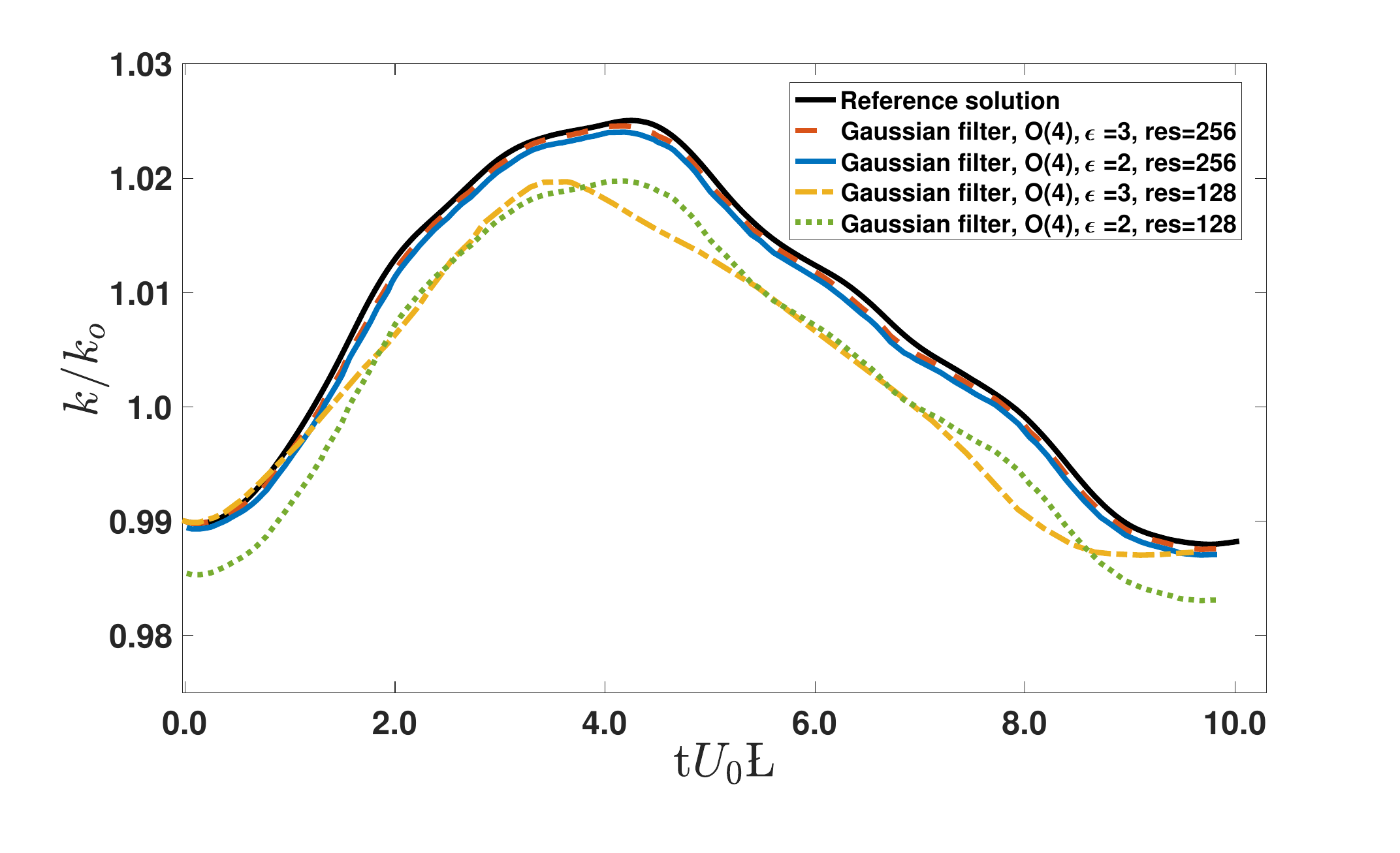}
           \caption{The dimensionless kinetic energy evolution in time, evaluated by the second and fourth order Gaussian filter with $\mathit{Re} = 4 \times 10^3$ by utilising the DC-PSE method at different resolutions. The results are compared to the DNS result form JHTDC~\cite{Li:2008b}. The fourth order filter with $res=256$ shows a strong alignment with the reference solution, at lower resolution there is a noticeable decline in the rate of kinetic energy dissipation.}
          \label{fig:k}
\end{figure}

In Fig.~\ref{fig:k} we present the evolution of dimensionless kinetic energy $k$ (here normalised by the initial kinetic energy $k_0$ reported in the reference study) 
using both filters with different computational resolution, compared with the total kinetic energy provided by the Johns Hopkins Turbulence Database (JHTDC), is presented in the referenced Fig.~\ref{fig:k}. JHTDC documents the kinetic energy evolution from time $t = 0$ to $t=10.04$, based on direct numerical simulation (DNS) data.

\begin{figure}[H]
\center
     \includegraphics[clip, width=\textwidth]{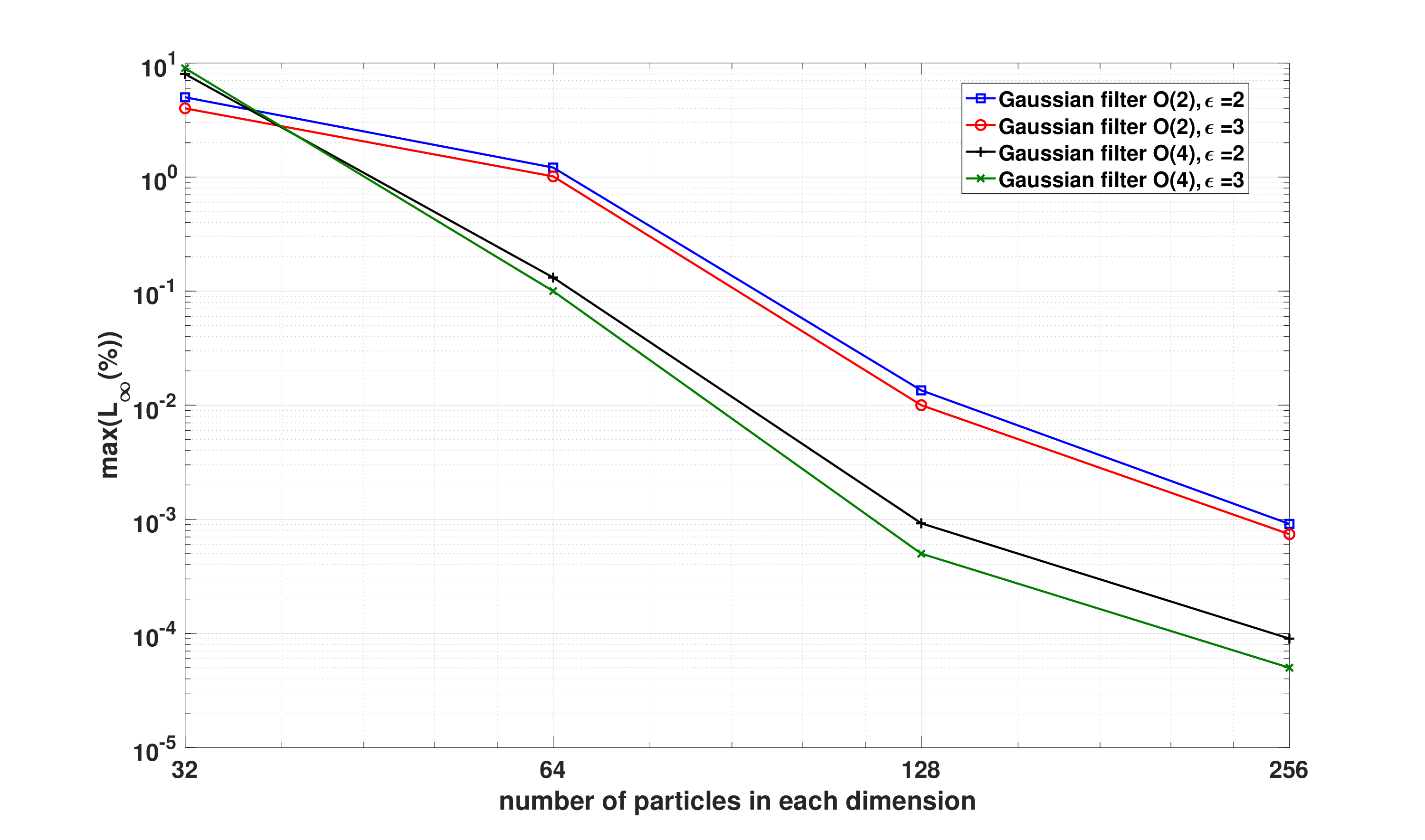}
           \caption{Demonstration of the $max(L_\infty(\%))$ error for the Gaussian filter of order $2$ and $4$, with different values for $\epsilon=2 $ and $3$. The filter solution's accuracy is significantly influenced by the spatial discretization.}
          \label{fig:lerror}
\end{figure}

The fourth-order Gaussian filter ($\textbf{G-O(4)}$)
with a resolution ($res=256$) demonstrates close alignment with the reference solution, showing a strong correlation in terms of kinetic energy dissipation. Furthermore, the parameter $\epsilon$ exhibits a minimal effect at this higher resolution. Conversely, at lower resolutions ($res=128$) , there is a noticeable decline in the rate of kinetic energy dissipation, with the disparity becoming more pronounced at $\epsilon =2$. This highlights the significant role both resolution and $\epsilon$ value play in achieving accurate representation of energy dissipation., which is later seen in Fig.~\ref{fig:lerror}.
In Fig.~\ref{fig:lerror}, we present the ($L_\infty$) error associated with the kinetic energy solution, underscoring the impact of the order of the filter, the ($\epsilon$) parameter and the resolution on solution precision. This error metric is derived by comparing the kinetic energy's time evolution from JHTDB against results achieved with various filter orders and \(\epsilon\) settings.

Observations from Fig.~\ref{fig:lerror} reveal that the filter solution's accuracy is significantly influenced by the spatial discretization; to be able to catch the flow dynamic filters of higher order necessitate a more refined particle distribution to reach the requisite accuracy. The fourth-order Gaussian filter demonstrates a greater $L_\infty$ error at lower resolutions in contrast to the second-order filter. However, at increased resolutions, the fourth-order filter closely approximates the DNS solution.

Moreover, the filter's performance is influenced by the selection of $\epsilon$, adjusting the point cloud's resolution concerning the intended filter's cutoff length scales. An $\epsilon$ value of $3$ is shown to improve solution accuracy beyond an $\epsilon$ of $2$, offering a more precise estimation of the effects of small scales by more efficiently excluding the smaller scales.

Finally we present the energy spectra calculated as $E(k)$:
\begin{equation}
   E(k)= 4 \pi k^2 \left \langle  E(\textbf{k}) \right \rangle,
   \label{spectra}
\end{equation}

where $\left \langle ... \right \rangle$ is an average over the thin spherical shells of radius radius $k = \left |  \mathbf{k} \right |$.
Here, for each component of the velocity fields $U=(u_i, u_j, u_z)$ Fourier transformation is as $\widehat{u}=(u_{ki}, u_{kj}, u_{kz})$, then the velocity spectrum tensor 
is computed as~\cite{Shi:2012}:
\begin{equation}
   E(\mathbf{k})=\frac{1}{2}\left | \widehat{u}(\mathbf{k})\cdot\widehat{u}^*(\mathbf{k}) \right |,
\label{eq:spectra}
\end{equation}
where $\widehat{u}^*$ is the complex conjugate of the transformed velocity, and
$\mathbf{k}=(k_i, k_j, k_z)$ is the wave number.
Finally the energy spectrum $E(k)$ is obtained as 

\begin{figure}[H]
\center
     \includegraphics[clip, width=0.8\textwidth]{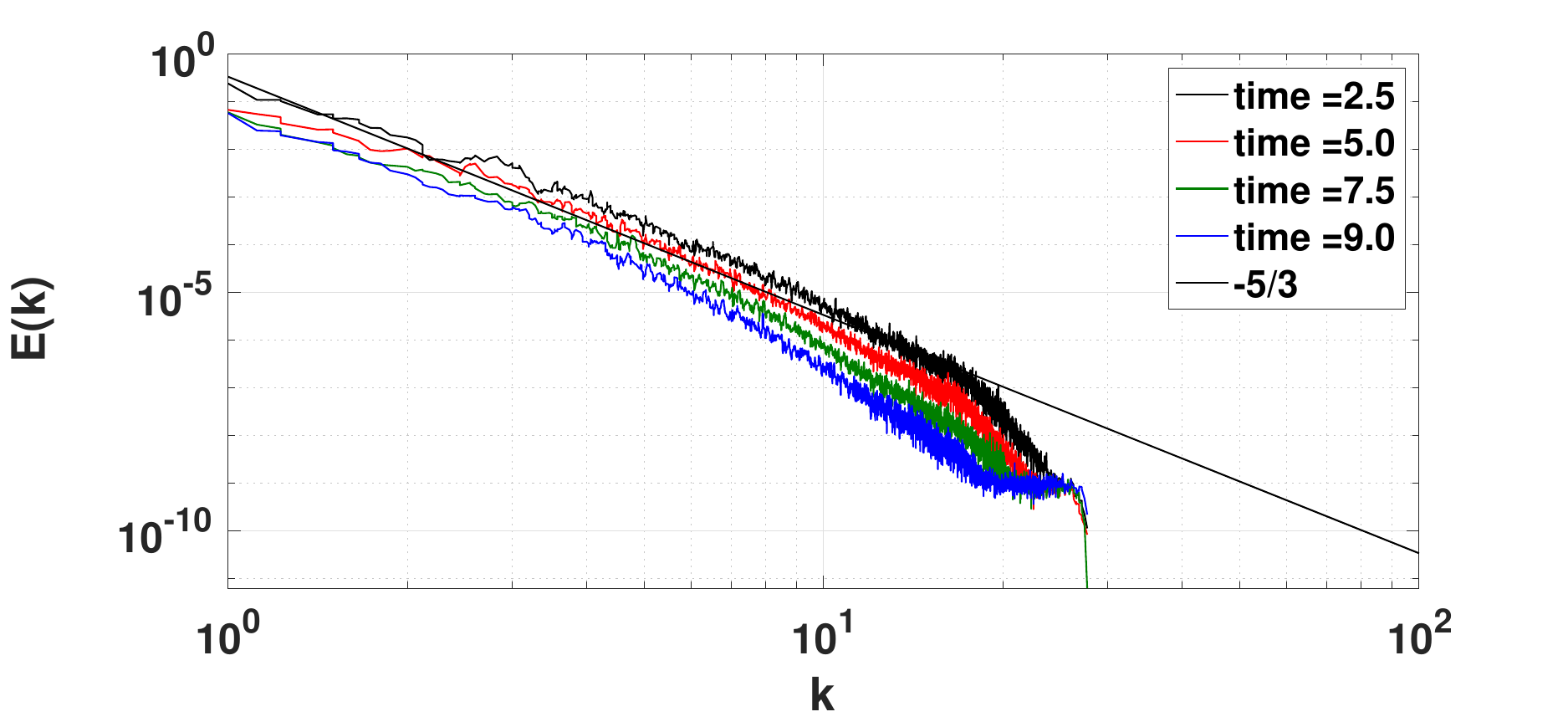}
           \caption{Time evolution of the energy spectra, evaluated at different time during the simulation, using the DC-PSE method with fourth order Gaussian filter. The energy dissipation is in good agreement with the Kolmogorov $-5/3$ profile.}
          \label{fig:spec}
\end{figure}

Fig.~\ref{fig:spec} represents the temporal energy spectrum evolution calculated as in Eq.~\ref{spectra} using the $\textbf{G-O(4)}$, $res =128$, at $\mathit{Re} = 4 \times 10^3$. This figure showcase the energy spectra for different states in dimensionless time $2.5,5.0,7.5 $ and $9.0$  we can observe that the energy dissipation is in good agreement with the Kolmogorov $-5/3$ profile.

\begin{figure}[H]
\center
     \includegraphics[clip, width=0.8\textwidth]{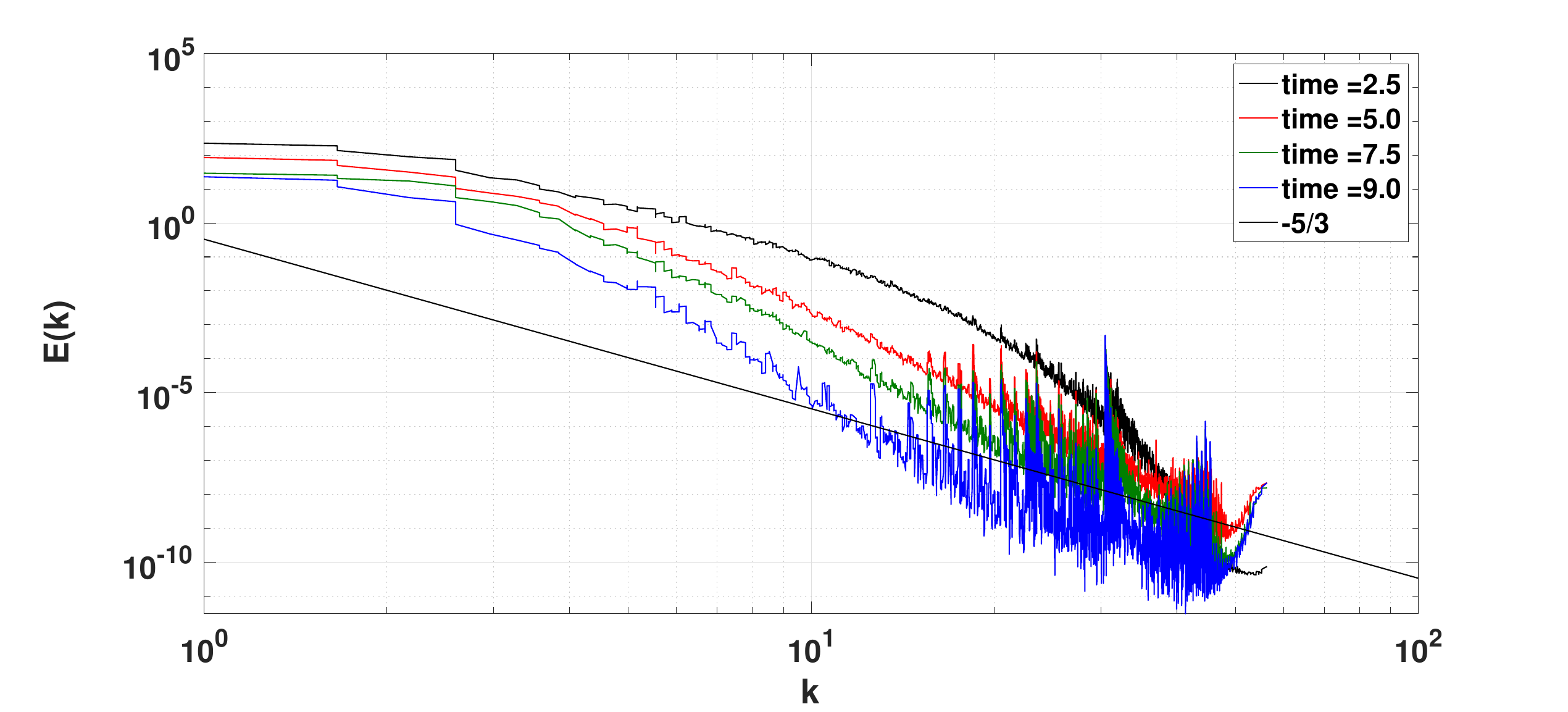}
           \caption{Time evolution of the energy spectra, evaluated at different time during the simulation, using the DC-PSE method with second order Gaussian filter, along with the Kolmogorov $-5/3$ profile. A noticeable noise within the
spectra becomes evident.}
          \label{fig:spec2}
\end{figure}

The $\textbf{G-O(4)}$ effectively capturing high-frequency behaviors, as seen in Fig.~\ref{fig:spec}.  Conversely, the application of the $\textbf{G-O(2)}$, presented in Fig.~\ref{fig:spec2}, resulted in less satisfactory results at high frequencies, where noticeable noise within the spectra becomes evident. 
The solver's inability to accurately represent high frequencies using 
$\textbf{G-O(2)}$ can likely be attributed to to the fact that  lower-order filters like are more susceptible to dispersion and dissipation errors, which can distort the energy spectrum, particularly at high frequencies. Resulting in a less accurate representation of the energy distribution across different scales.

\begin{figure}[H]
  \begin{subfigure}[b]{0.5\textwidth}
  \centering
     \includegraphics[trim = {1cm 0 1cm 0},clip, width=0.85\textwidth]{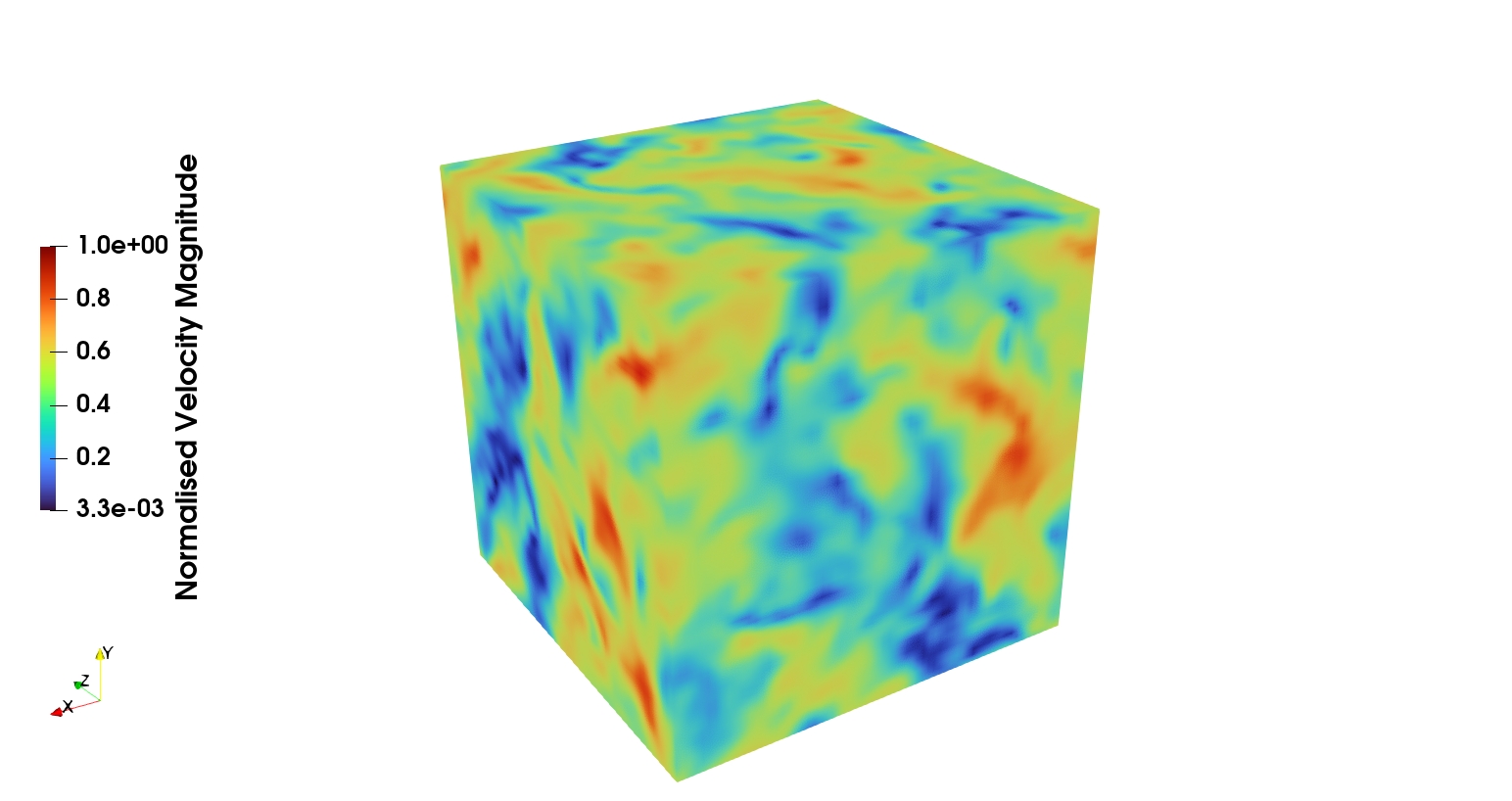}
          \caption{time=0.0}
       \end{subfigure}
       \hfill
        \begin{subfigure}[b]{0.5\textwidth}               
       \centering
       \includegraphics[trim = {1cm 0 1cm 0}, clip, width=0.85\textwidth]{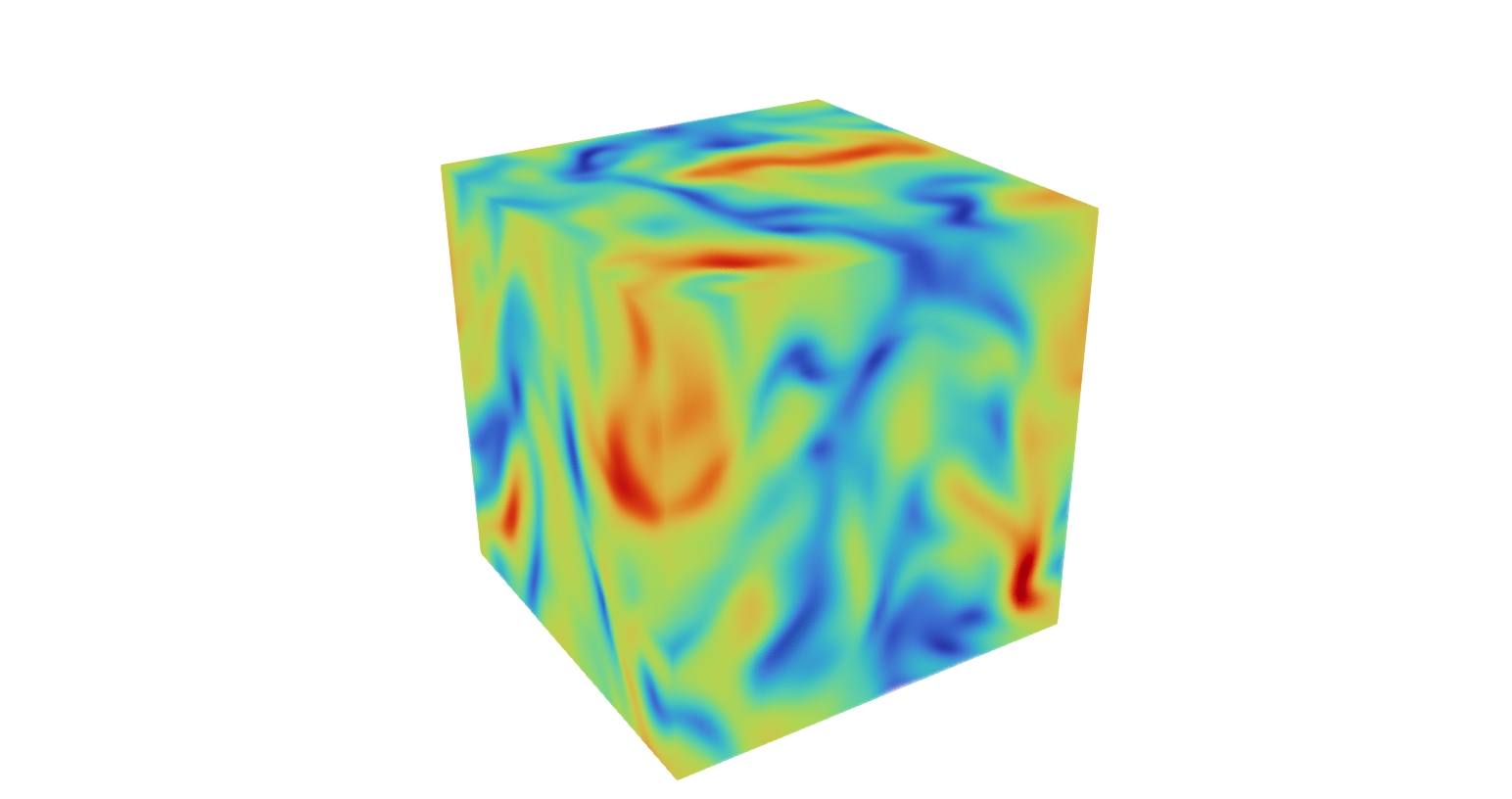}
                   \caption{time=2.5}
             \end{subfigure}
      \begin{subfigure}[b]{0.5\textwidth}
        \centering      
       \includegraphics[trim = {1cm 0 1cm 0},clip, width=0.85\textwidth]{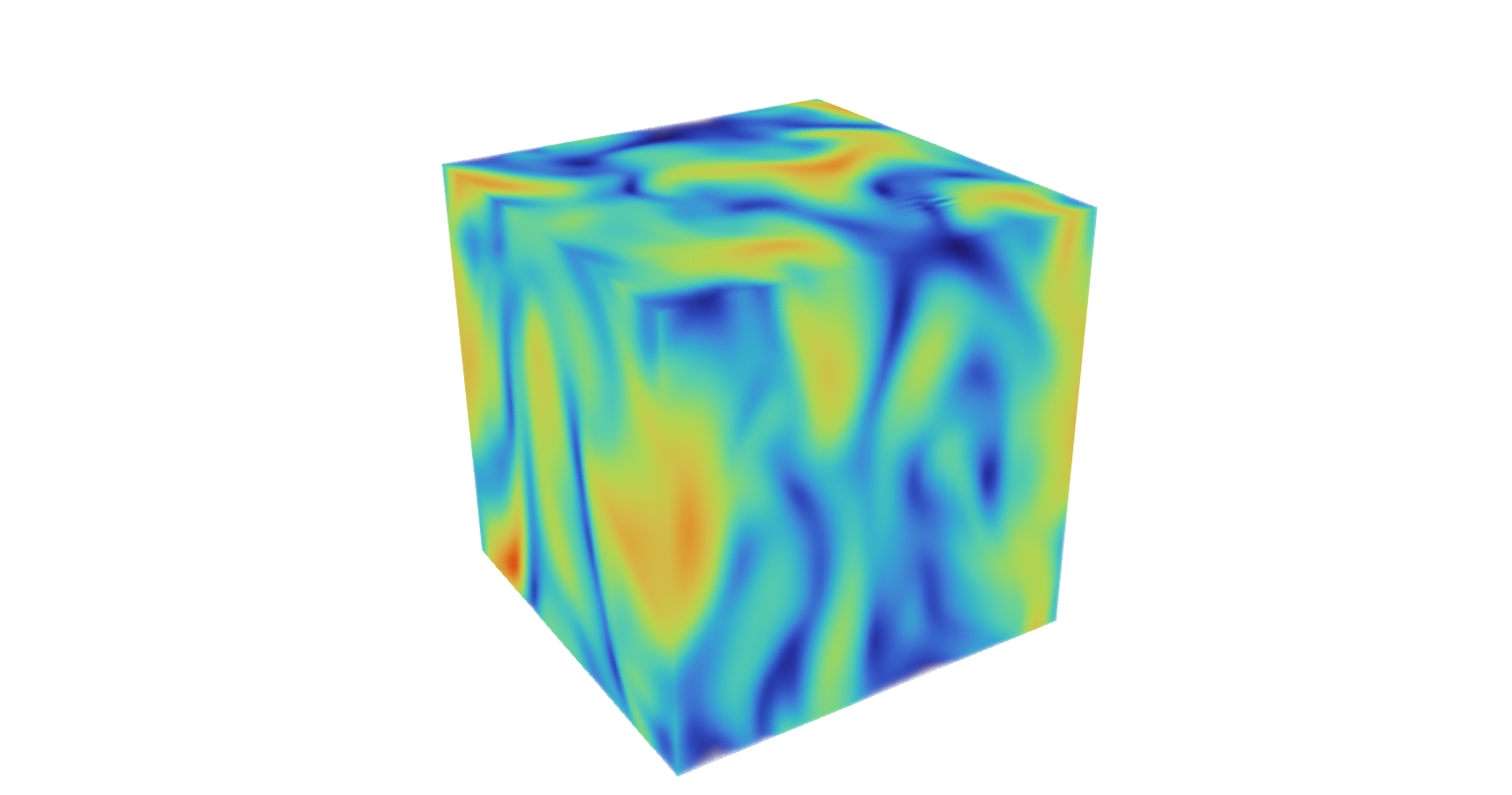}
             \caption{time=5.0}
             \end{subfigure}
             \begin{subfigure}[b]{0.5\textwidth}
             \centering
       \includegraphics[trim = {1cm 0 1cm 0},clip, width=0.85\textwidth]{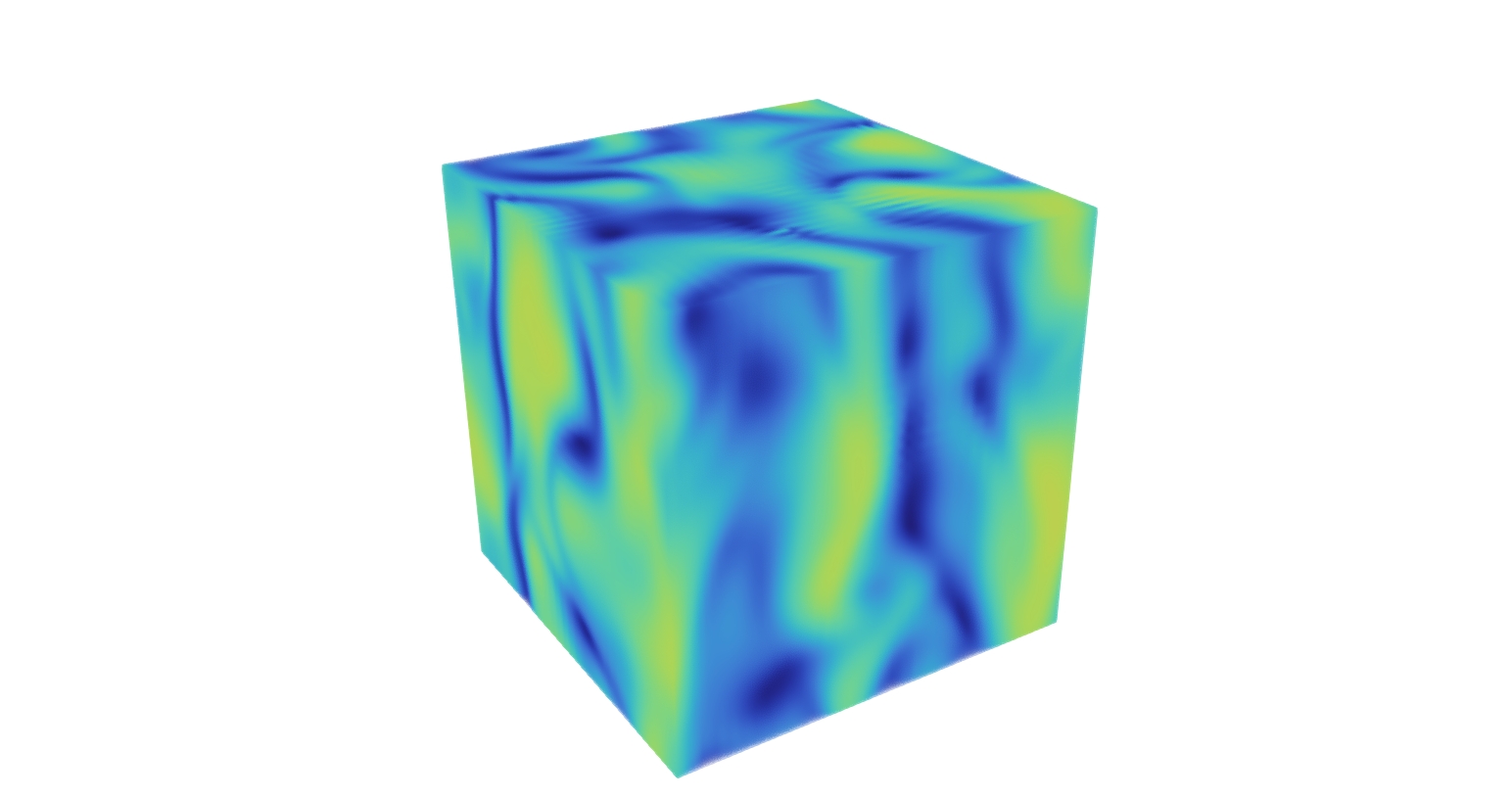}
                                           \caption{time=7.5}
             \end{subfigure}
  \caption{(a)-(d) show the decay of the non-dimensionalised velocity evolution $(U/Max(U_{o}))$ for the three dimensional isotropic turbulence case in time. The method captures the development of the anticipated large-scale structure successfully.}
          \label{fig:U3d}
\end{figure}

\begin{figure}[H]
  \begin{subfigure}[b]{0.5\textwidth}
  \centering
     \includegraphics[trim = {3cm 0 3cm 0},clip, width=0.85\textwidth]{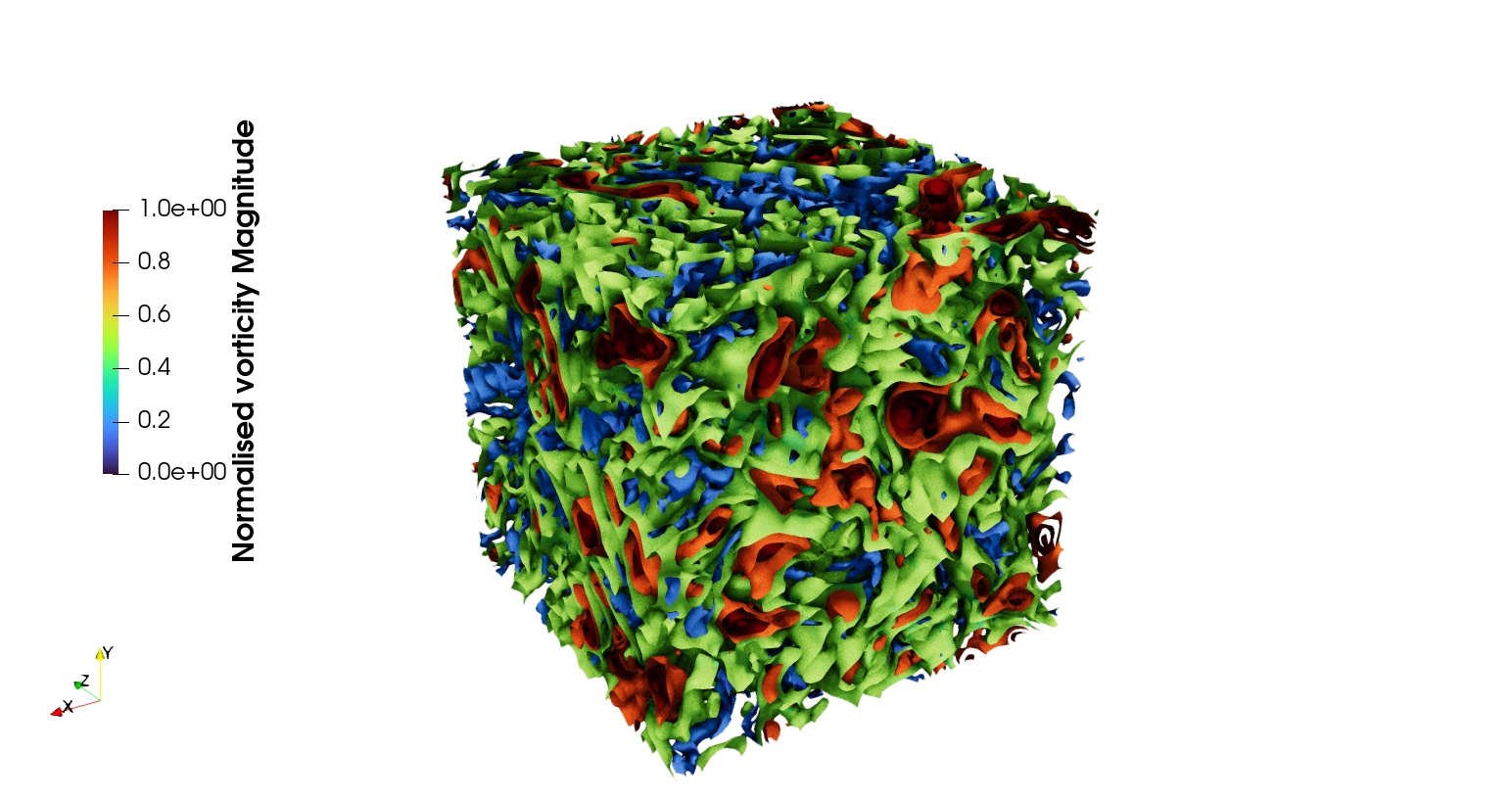}
          \caption{time=0.0}
       \end{subfigure}
       \hfill
        \begin{subfigure}[b]{0.5\textwidth}               
       \centering
       \includegraphics[trim = {3cm 0 3cm 0}, clip, width=0.85\textwidth]{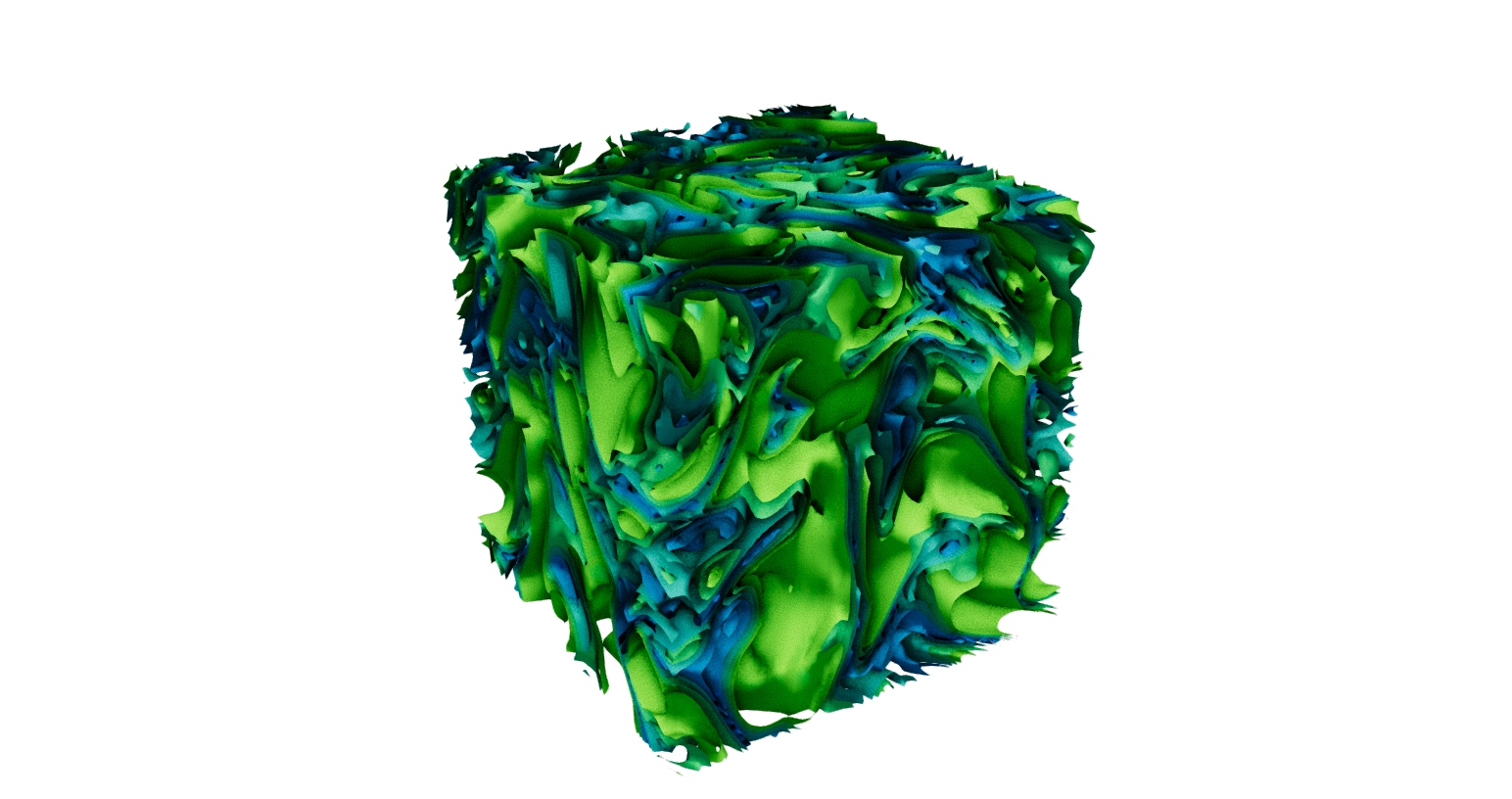}
                   \caption{time=2.5}
             \end{subfigure}
      \begin{subfigure}[b]{0.5\textwidth}
        \centering      
       \includegraphics[trim = {3cm 0 3cm 0},clip, width=0.85\textwidth]{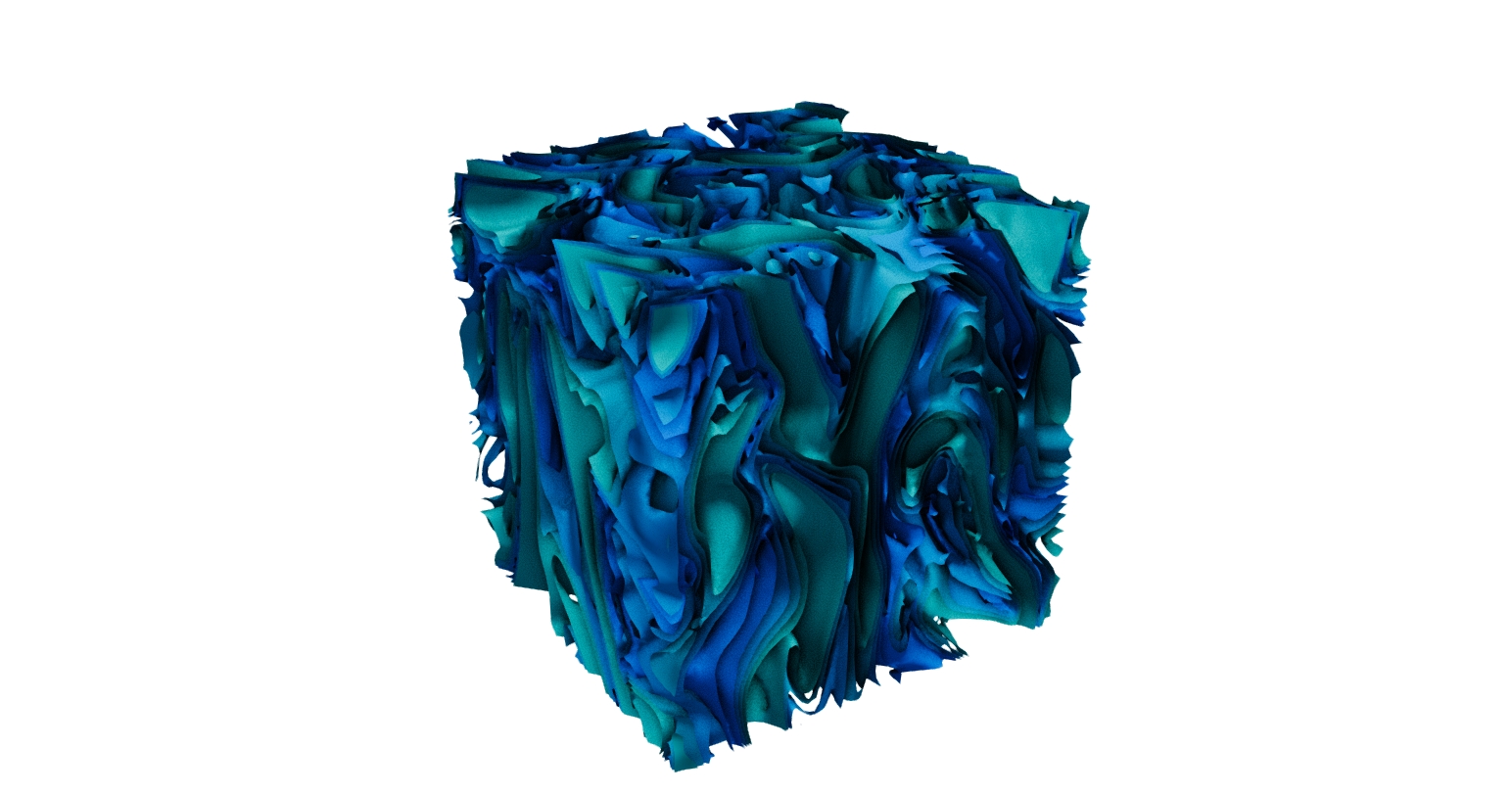}
             \caption{time=5.0}
             \end{subfigure}
             \begin{subfigure}[b]{0.5\textwidth}
             \centering
       \includegraphics[trim = {3cm 0 3cm 0},clip, width=0.85\textwidth]{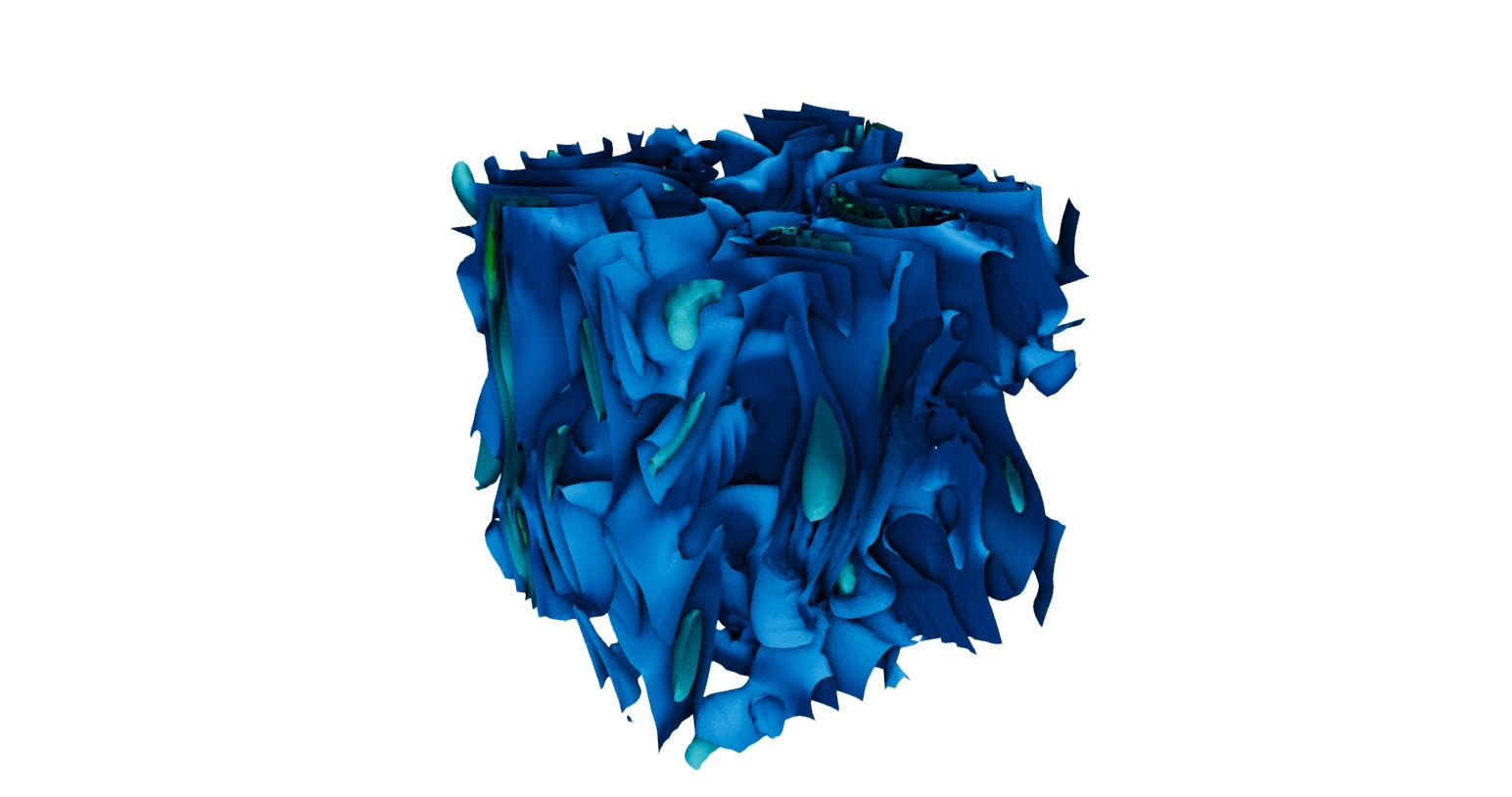}
        \caption{time=7.5}
             \end{subfigure}
  \caption{(a)-(d) show the decay of the non-dimensionalised vorticity evolution $(\omega/Max(\omega_{o}))$ for the three dimensional isotropic turbulence case in time.}
          \label{fig:V3d}
\end{figure}

The evolution of the flow is depicted in Figures~\ref{fig:U3d}(a-d) and \ref{fig:V3d}(a-d), showcasing the non-dimensionalised velocity $(U/Max(U_{o}))$ and vorticity $(\omega/Max(\omega_{o}))$ fields. Here $Max(U_{o})$ and $Max(\omega_{o})$ are the maximum initial velocity and vorticity. The evolution time is over a non-dimensionalised time period from $t = 0$-$9$. The introduced method successfully captures the development of the anticipated large-scale structures, as illustrated in Figure~\ref{fig:V3d}(a-d), highlighting the method's capability in capturing the dynamic evolution of the flow's key features.

\subsection{An impulsively started three-dimensional flow surrounds four cylinders arranged in a square in-line configuration} 
\label{flow-cylinder}
A cylinder is one of the basic shapes of any structural engineering component, and often used in groups. Conducting numerical analyses on the fluid flow and vortex shedding around a simple cylinder setup is essential in understanding the flow around more intricate structures. It also acts as a critical benchmark for verifying the current the presented approach.

\begin{figure}[H]
\center
     \includegraphics[clip, width=0.7\textwidth]{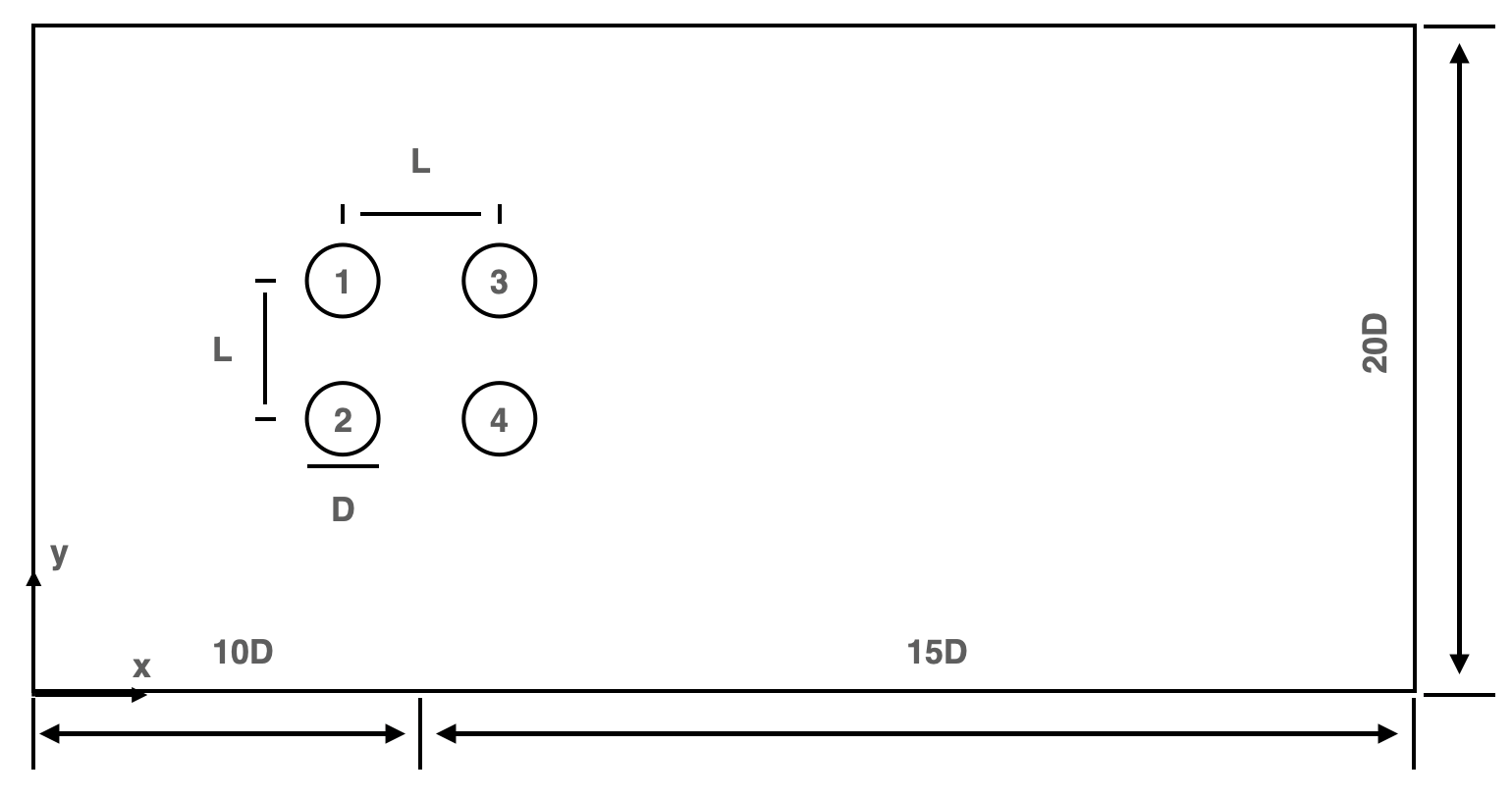}
           \caption{Schematic representation of the computational domain of the four cylinders arranged in a square in-line configuration.}
          \label{fig:cyldomain}
\end{figure}

In this benchmark, we explore an impulsively initiated flow around a group of four cylinders arranged in an in-line square configuration. A two-dimensional (2-D) schematic of the three-dimensional (3-D) computational domain is depicted in Fig.~\ref{fig:cyldomain}, where $D$ represents the cylinder diameter and $L$
 denotes the inter-cylinder distance. The cylinders are positioned in an array with spacing ratios of $L/D=1.5$ and $3.5$.
 
The flow is characterised by the following non-dimensional quantities, Reynolds number $\mathit{Re}$, time $t$ and Mach number $Ma$, defined by the equations: ,

\begin{equation}
\mathit{Re}=U_\infty \rho D/\mu,
\qquad
t=\delta t U_\infty/D,
\qquad
Ma=Uo/cs 
\end{equation}

Where $U_\infty$ denotes the free-stream velocity of the flow.
For this test case, the flow around a cylinder array is investigated at $\mathit{Re} = 1.5 \times 10^4$ and a Mach number $Ma = 0.15$. The flow velocity component is uniformly distributed as $U=(u=U_o, v=0, w=0)$, alongside a uniform distribution of pressure 
$p_o$ uniform pressure and density $\rho_0$ distribution, with $U_o$ is the reference velocity.

The computational domain surrounding the cylinder is bounded within $0<x<25D, 0<y<20 D$ and $0<z<5D$ with spacial spacing of $128$ particle in each coordinate axis
and applies periodic boundary conditions. No-slip boundary conditions are imposed on the cylinder surfaces utilising the penalization technique outlined in Eq.~(\ref{momentum-pen}), setting $u_{oq} =0$.

\begin{figure}[H]
\center
     \includegraphics[clip, width=0.8\textwidth]{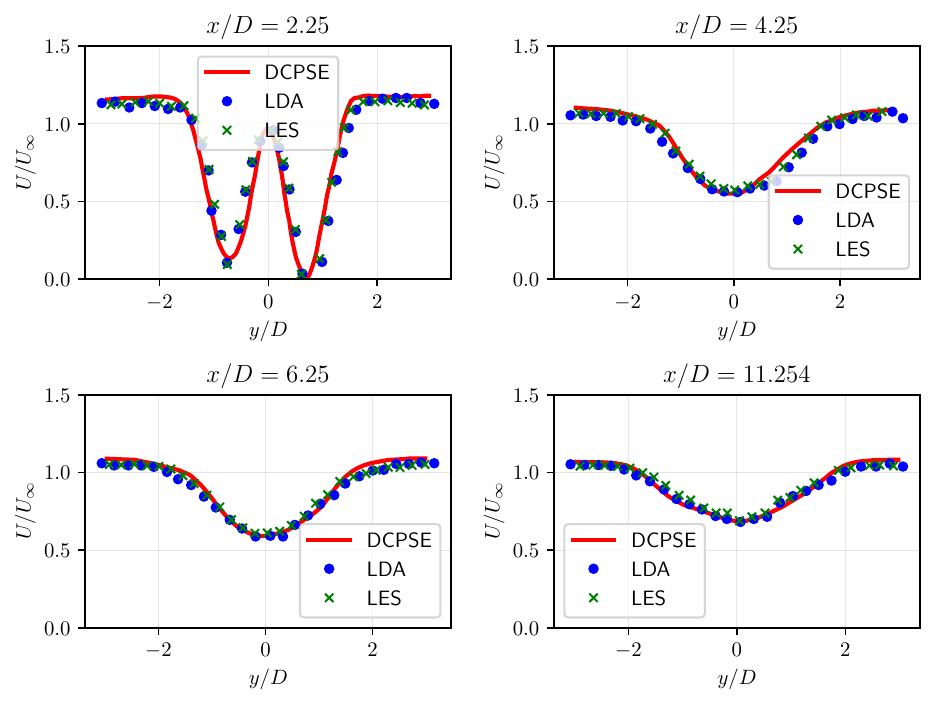}
           \caption{The mean streamwise velocity distribution profiles at a dimensionless distance $x/D$, compared with the LDA measurements and LES simulation presented in ~\cite{Lam:2009} for the four cylinders in inline square configuration with spacing ration $L/D=1.5$. At $x/D= 2.25$ a wake like profile indicating a reversal flow in the region behind the third cylinder, going downstream the wake widen indicating the formation of a combined structure.}
          \label{fig:cyl15}
\end{figure}

\begin{figure}[H]
\center
     \includegraphics[clip, width=0.7\textwidth]{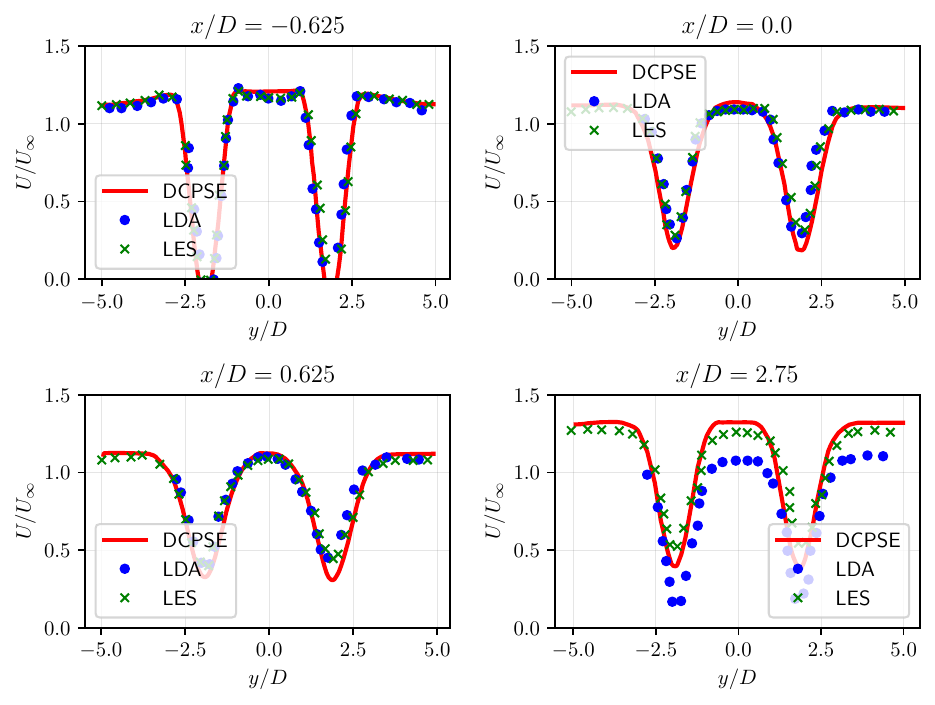}
           \caption{The mean streamwise velocity distribution profiles at a dimensionless distance $x/D$, compared with the LDA measurements and LES simulation presented in ~\cite{Lam:2009} for the four cylinders in inline square configuration with spacing ration $L/D=3.5$. A reverse flow behind cylinders $1$ and $2$ is observed at $x/D= -.625$ and vortex shedding from the upstream cylinders is initiated. Moving downstream we do not observe the same combined wake structure as in~\ref{fig:cyl15}}
          \label{fig:cyl35}
\end{figure}

In Figs.~\ref{fig:cyl15} and~\ref{fig:cyl35}, we display the normalised mean streamwise velocity distributions $U/U_{\infty}$ in the near wake region of the cylinders, spacing ratios of $L/D = 1.5$ and $3.5$, respectively. The profiles of streamwise velocity are determined at a dimensionless distance $x/D$ from the centre of the four-cylinder configuration and are compared with Laser Doppler Anemometry (LDA) measurements and LES reported in~\cite{Lam:2009}.

For the spacing ratio $L/D=1.5$, the mean streamwise velocity profiles display a slightly asymmetrical wake-like profile at the position $x/D=2.25$, implying the presence of reverse flow in the region behind the third cylinder, as shown in Fig.~\ref{fig:cyl15}. As the flow progresses downstream from $x/D=4.25$ to $x/D=11.25$, the wake-like profile widens, and the normalised mean streamwise velocity profiles indicate the formation of a combined structure, resulting in a singular, expansive wake.

Adjusting the spacing ratio $L/D$ from $1.5$ to $3.5$ has a noticeable effect on the shape of the mean streamwise velocity profiles, as seen in Fig~\ref{fig:cyl35}.
First, at $x/D= -0.625$ the $U/U_{\infty}$ values reach negative values at $y/D \approx 2.0, -2.0$, indicating a reverse flow behind cylinders $1$ and $2$ and vortex shedding from the upstream cylinders is initiated. Moving downstream from $x/D= 0.0$ to $x/D= 2.75$, there is an apparent reduction in the length over which vortex formation occurs. Furthermore, the wake structure behind the downstream cylinders ($3$ and $4$) maintains symmetry, diverging from the combined wake structure observed at the smaller spacing ratio.

\begin{figure}[H]
\center
     \includegraphics[clip, width=0.7\textwidth]{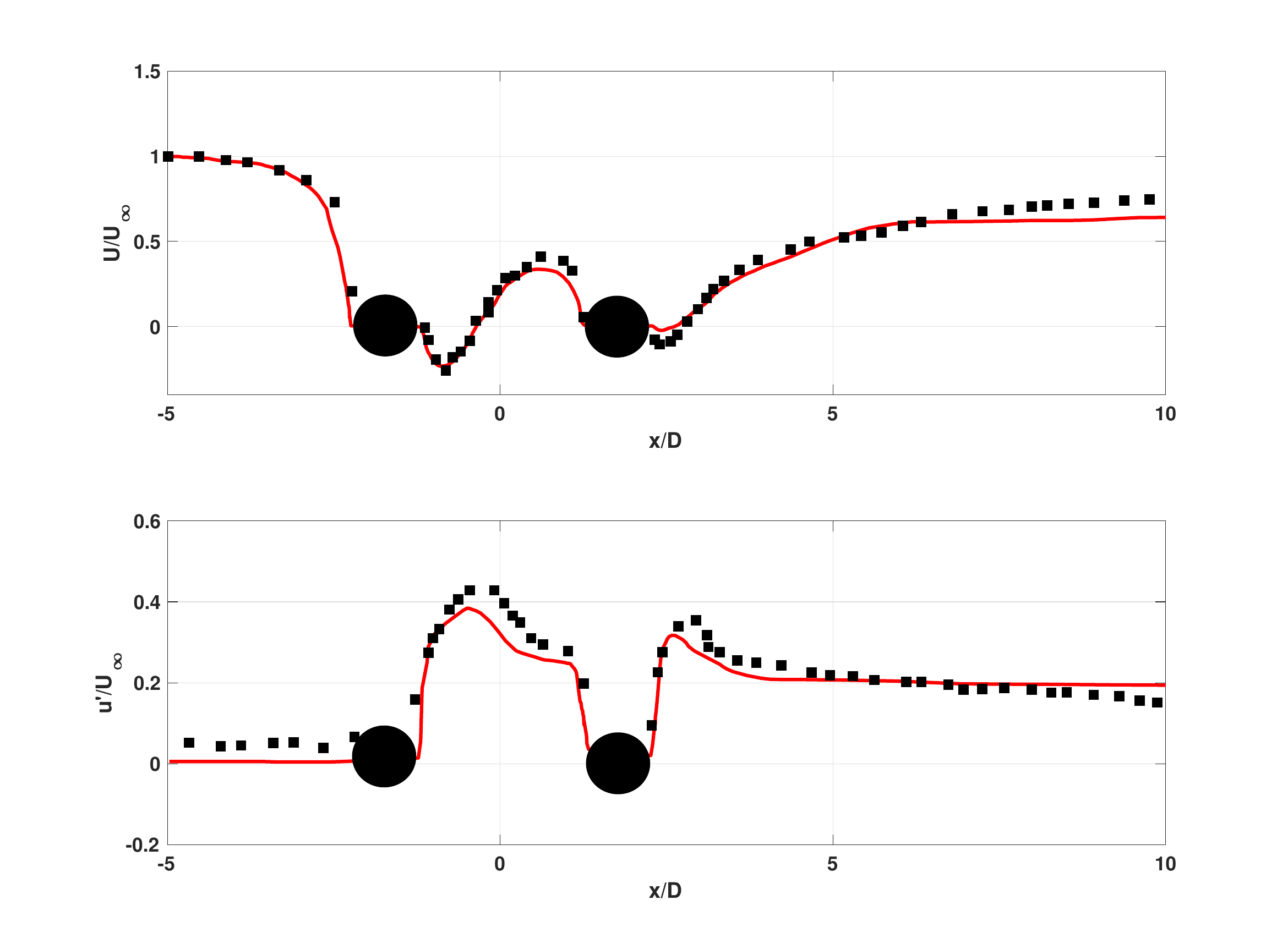}
           \caption{(a) The mean stream velocity $(U/U_{\infty})$, (b) fluctuation streamwise velocity $(u'/U_{\infty})$. Both profiles are compared with the LDA measurements presented in ~\cite{Lam:2009} along the $y$-plane ($y/D =1.75$) going through cylinder $2$ and $4$. The numerical method has a good agreement with the LDA measurements for both profiles. However, moving downstream the DC-PSE predicts a lower $(U/U_{\infty})$ and a higher $(u'/U_{\infty})$, between the two cylinders we observe a shorter vortex formation from the DC-PSE simulation compared to the LDA measurements. }
          \label{fig:cylU}
\end{figure}

Figure~\ref{fig:cylU} illustrates the distribution of normalised mean streamwise velocity $(U/U_{\infty})$ and normalised root mean square (RMS) values of fluctuating streamwise velocity $(u'/U_{\infty})$ along a centre-line plane downstream of tandem cylinders $2$ and $4$ with a spacing ratio $L/D = 3.5$. Comparing our findings with the LDA data reported in~\cite{Lam:2009}, we observe close agreement between the numerical results and the LDA measurements for both the normalised mean streamwise velocity $(U//U_{\infty})$ distribution and the normalised RMS values of fluctuating streamwise velocity $(u'/U_{\infty})$. However, the DC-PSE model predicts a lower $(U/U_{\infty})$ and a higher $(u'/U_{\infty})$ downstream of the computational domain. Additionally, in the region between the two cylinders, the DC-PSE model exhibits shorter vortex formation compared to the LDA measurements.

In Table~\ref{table}, we present the root mean square of the fluctuating lift coefficients ($\overline{C}'{Li}$) and the Strouhal numbers ($St_i$), where the subscript ($i$) indicates the cylinder number (ranging from $1-4$), alongside comparisons to the reference solution. Significant findings can be observed in the wake length behind the downstream cylinders, as evidenced by the variation in $\overline{C}'{L4}$ and $\overline{C}'{L3}$.


\begin{table}[]
\center
\begin{tabular}{@{}lllllllllll@{}}
\toprule
\textbf{Study}    & \textbf{L/D} & \textbf{$\mathit{Re}$} & \textbf{$\overline{C}'_{L1}$} & \textbf{$\overline{C}'_{L2}$} & \textbf{$\overline{C}'_{L3}$} & \textbf{$\overline{C}'_{L4}$} & \textbf{$St_1$} & \textbf{$St_2$} & \textbf{$St_3$} & \textbf{$St_4$} \\ \midrule
\textbf{Reference solution} LES~\cite{Lam:2009}     &              & $1.5 \times 10^4$      &              &              &              &              &             &             &             &             \\
                   & 3.5          &             & 0.66        & 0.62         & 1.22         & 1.25         & 0.196       & 0.192       & 0.192       & 0.196       \\
                   & 1.5          &             & 0.10         & 0.06         & 0.28         & 0.34         &             &             & 0.125       & 0.169       \\
                   &              &             &              &              &              &              &             &             &             &             \\
\textbf{Present study} &              & $1.5 \times 10^4$       &              &              &              &              &             &             &             &             \\
                   & 3.5          &             & 0.62         & 0.62         & 1.24         & 1.25         & 0.196       & 0.196       & 0.194       & 0.194       \\
                   & 1.5          &             & 0.07         & 0.01         & 0.28         & 0.35         &             &             & 0.160       & 0.125       \\ \bottomrule
\end{tabular}

\caption{Computational results on mean fluctuating lift coefficients and Strouhal number of four cylinders in an in-line square configuration.}
\label{table}
\end{table}

\subsubsection{Flow structure}
\begin{figure}[H]
  \begin{subfigure}[b]{0.5\textwidth}
  \centering
     \includegraphics[trim ={0cm 11cm 0cm 0cm},clip, width=1.2\textwidth]{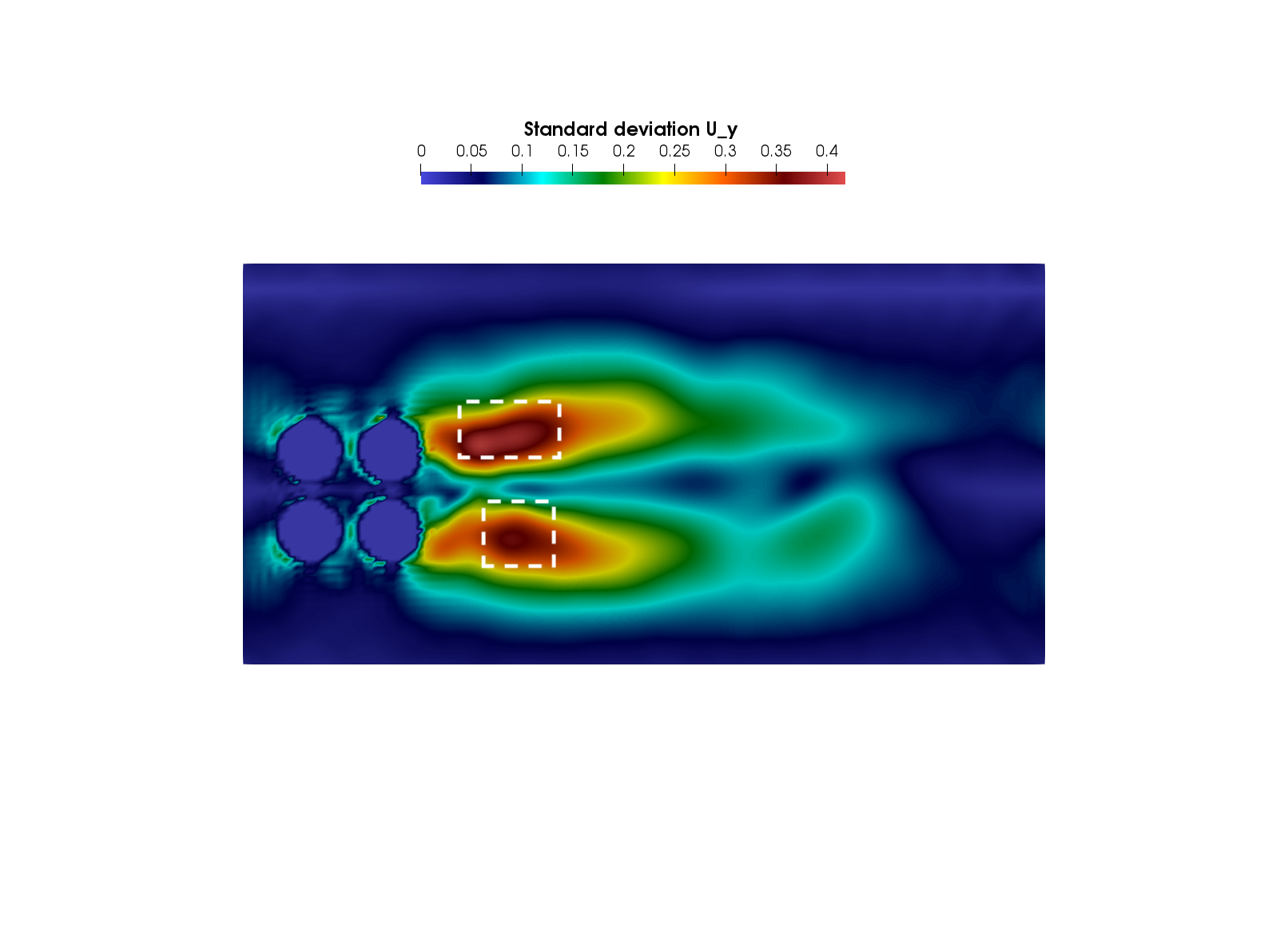}
     \caption{$L/D=1.5$}
       \end{subfigure}
  \begin{subfigure}[b]{0.5\textwidth}
  \centering
       \includegraphics[trim ={0cm 11cm 0cm 0cm},clip, width=1.2\textwidth]{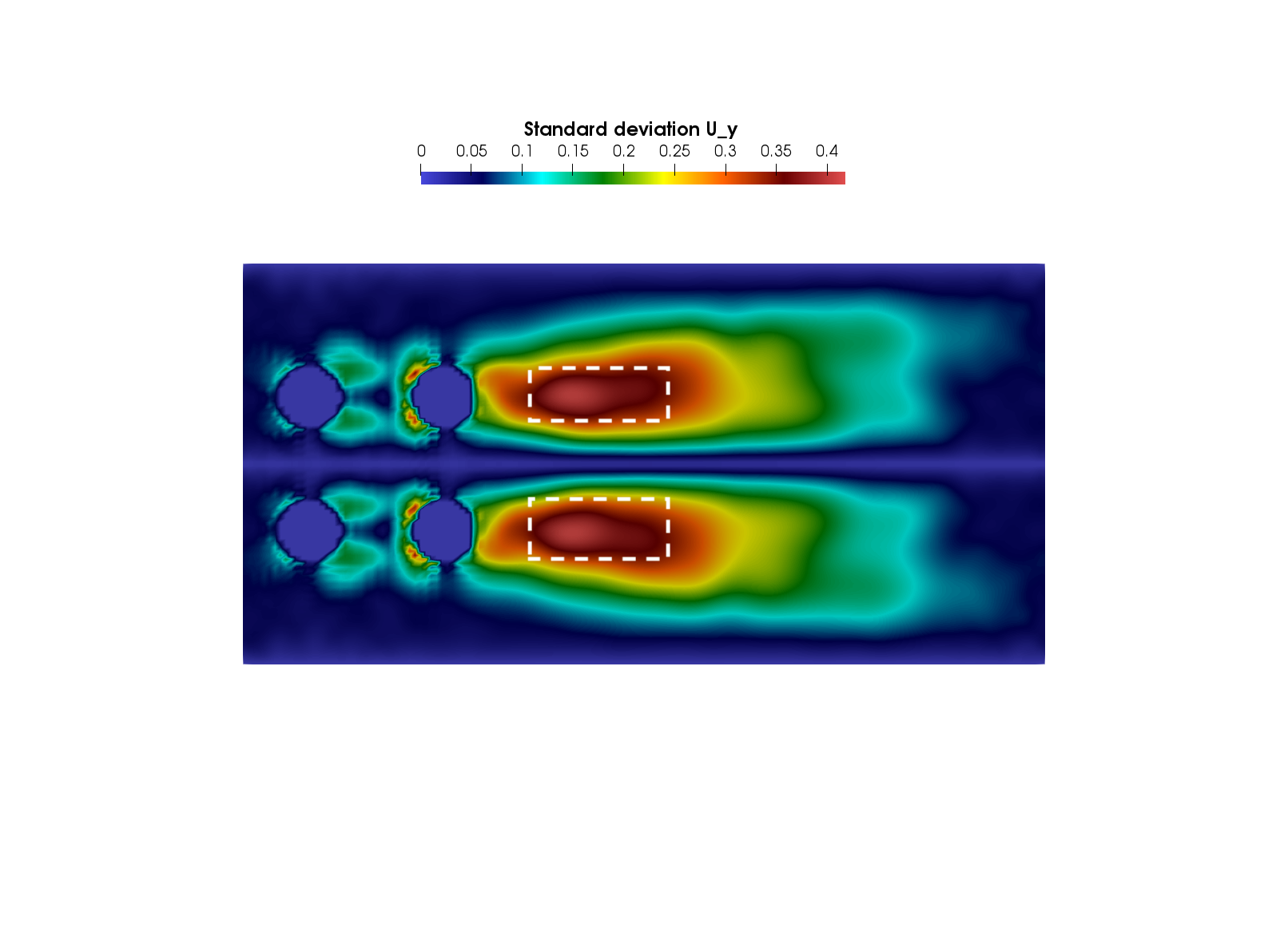}
            \caption{$L/D=3.5$}
             \end{subfigure}
          \caption{A cross section in the $z$-plane showing the standard deviation of the normalised velocity component $U_y$ of the flow for (a) spacing ratio $L/D=1.5$ and (b) $L/D=3.5$, where the region with intense fluctuation is marked with the dashed line rectangle. The fluctuation region is less intense and wider for $L/D=3.5$ compared to $L/D=1.5$.}
          \label{fig:std}
\end{figure}

The temporal variation in the flow field is calculated by the standard deviation of the velocity vector component ($U_y$), over time as presented as illustrated in Fig.~\ref{fig:std}.
For instance, with a spacing ratio of $L/D = 3.5$, the fluctuation region is more expansive and exhibits less intensity (denoted by a dashed rectangle) compared to  $L/D = 1.5$ ratio. 

\begin{figure}[H]
  \begin{subfigure}[b]{0.5\textwidth}
  \centering
     \includegraphics[trim ={0cm 15cm 0cm 7cm},clip, width=1.2\textwidth]{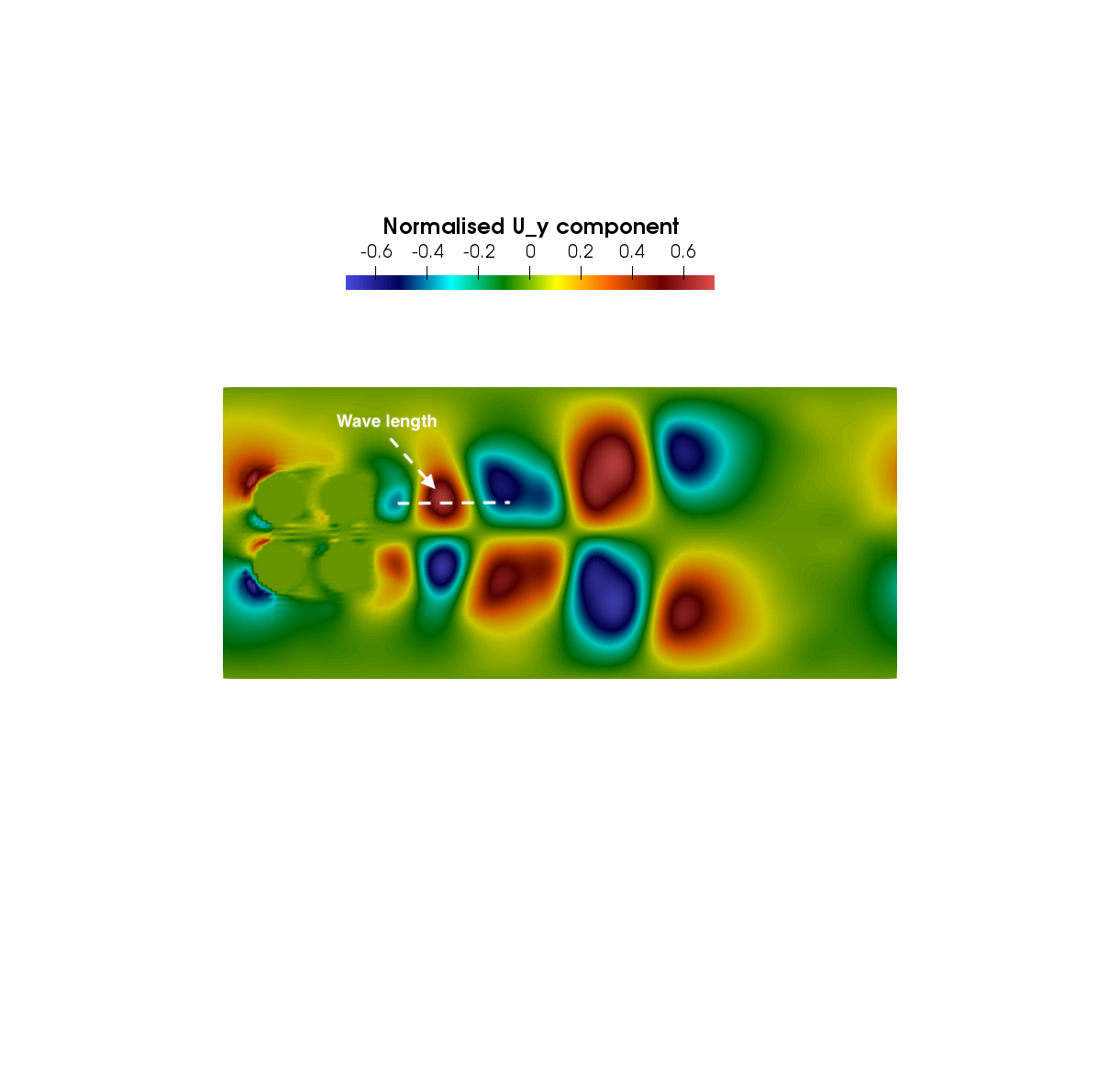}
     \caption{$L/D=1.5$}
       \end{subfigure}
  \begin{subfigure}[b]{0.5\textwidth}
  \centering
       \includegraphics[trim ={0cm 15cm 0cm 7cm},clip, width=1.15\textwidth]{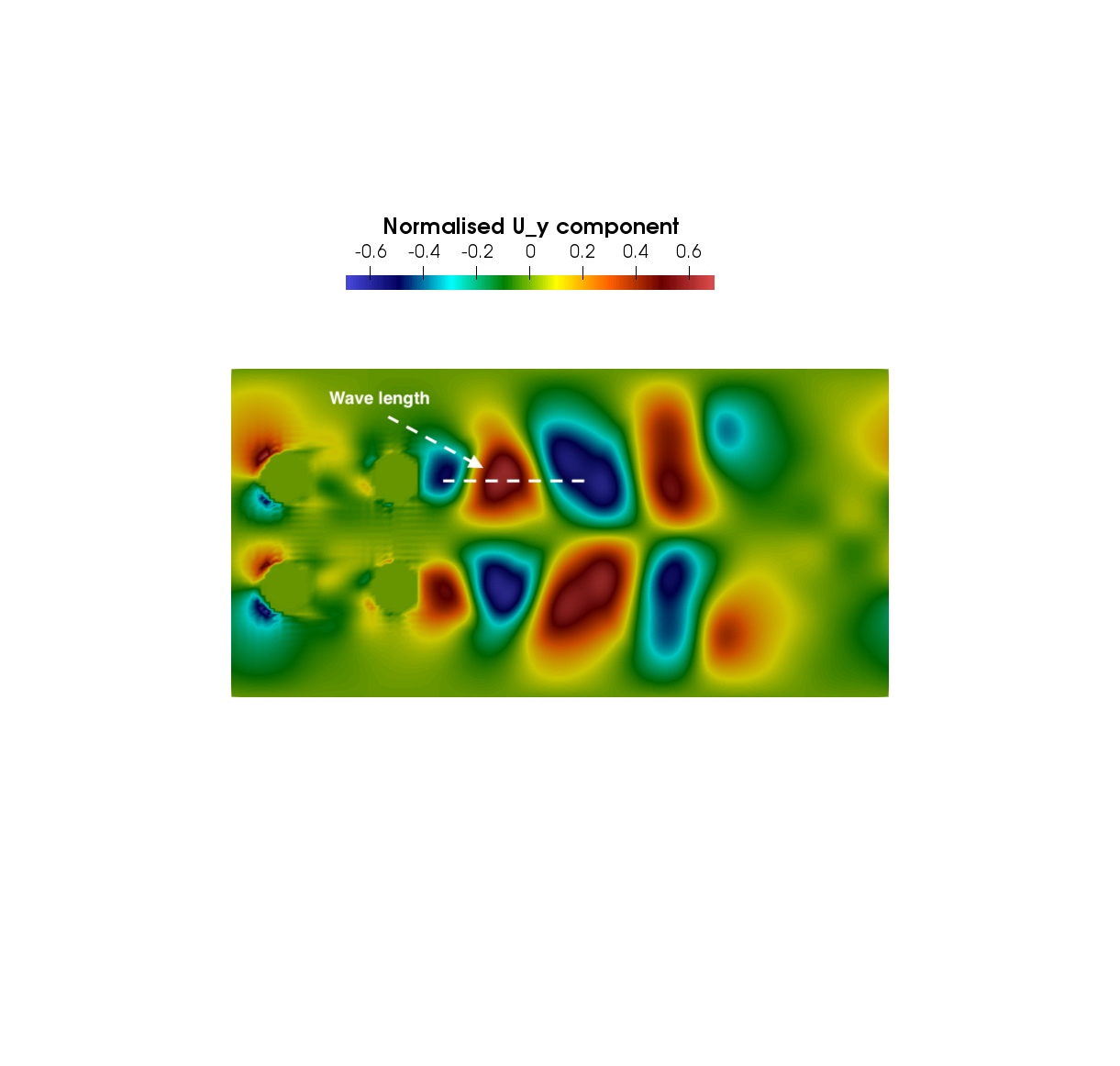}
       \caption{$L/D=3.5$}
             \end{subfigure}
          \caption{A cross section in the $z$-plane showing the normalised velocity component $U_y$ of the flow for (a) spacing ratio $L/D=1.5$ and (b) $L/D=3.5$, with the dashed line representing the wave length indicating the vortices spacing. The intensity of the vortices weakens with the enlargement of cylinder spacing.}
          \label{fig:wake}
\end{figure}

In Fig.~\ref{fig:wake} we present the normalised mean velocity vector component ($U_y$). As $L/D$ ratio becomes larger the vortices became spaced further apart,  the dashed line represents the wave length. Typically, the intensity of the vortices weakens with the enlargement of cylinder spacing.

\begin{figure}[H]
  \begin{subfigure}[b]{0.5\textwidth}
     \centering
     \includegraphics[trim ={0cm 10cm 0cm 7cm},clip, width=\textwidth]{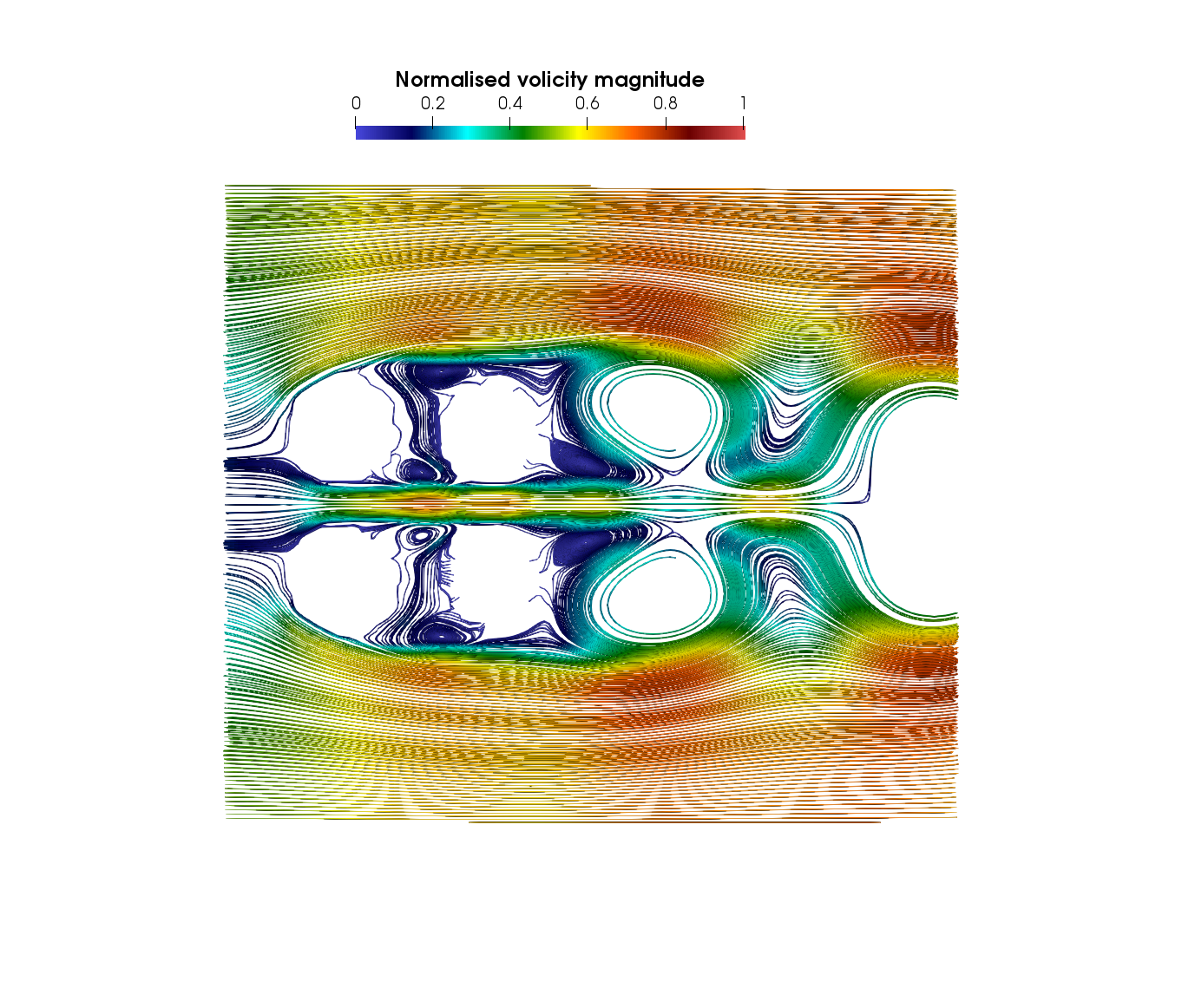}
     \caption{$L/D=1.5$}
    \end{subfigure}
  \begin{subfigure}[b]{0.5\textwidth}
     \centering
       \includegraphics[trim ={0cm 10cm 0cm 7cm},clip, width=\textwidth]{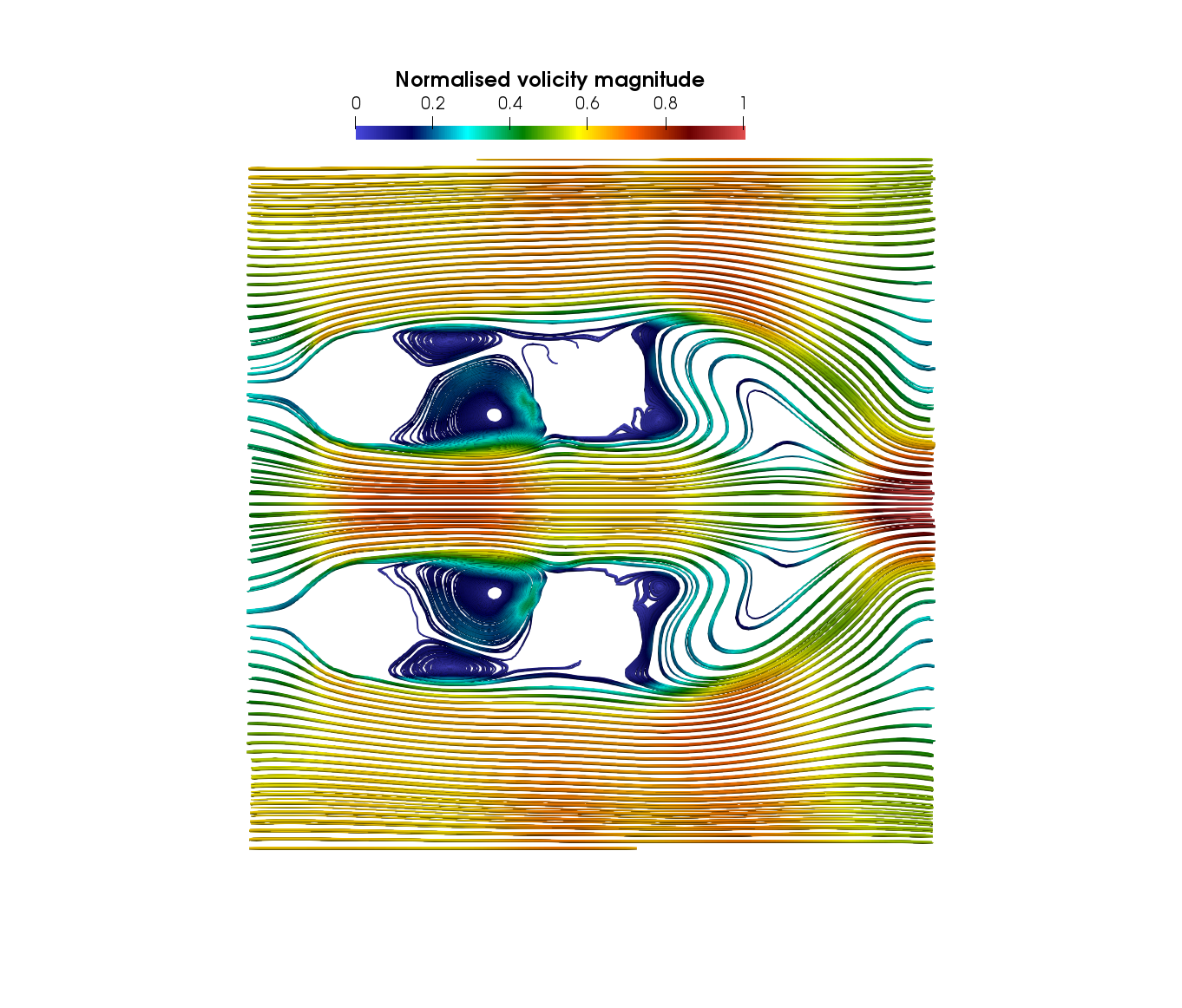}
       \caption{$L/D=3.5$}
   \end{subfigure}
   \caption{A cross section in the $z$-plane showing the streamlines of the normalised velocity of the flow for (a) spacing ratio $L/D=1.5$ and (b) $L/D=3.5$. At (a) with $L/D=1.5$ a large vortex emerges in the wake of the downstream, where at (b) $L/D=3.5$ we observe a large vortex formation behind the upstream cylinders $1$ and $2$.} 
          \label{fig:stream}
\end{figure}

In Fig.\ref{fig:stream} (b), a larger vortex forms behind the upstream cylinders ($1$ and $2$) compared to Fig.~\ref{fig:stream} (a), thereby affecting the downstream cylinders and the resultant wake formation. This distinction is less evident for the spacing ratio $L/D = 1.5$, where a larger vortex emerges in the wake of the downstream cylinders.

\begin{figure}[H]
\center
     \includegraphics[trim ={6cm, 11cm, 6cm 5cm},clip, width=0.7\textwidth]{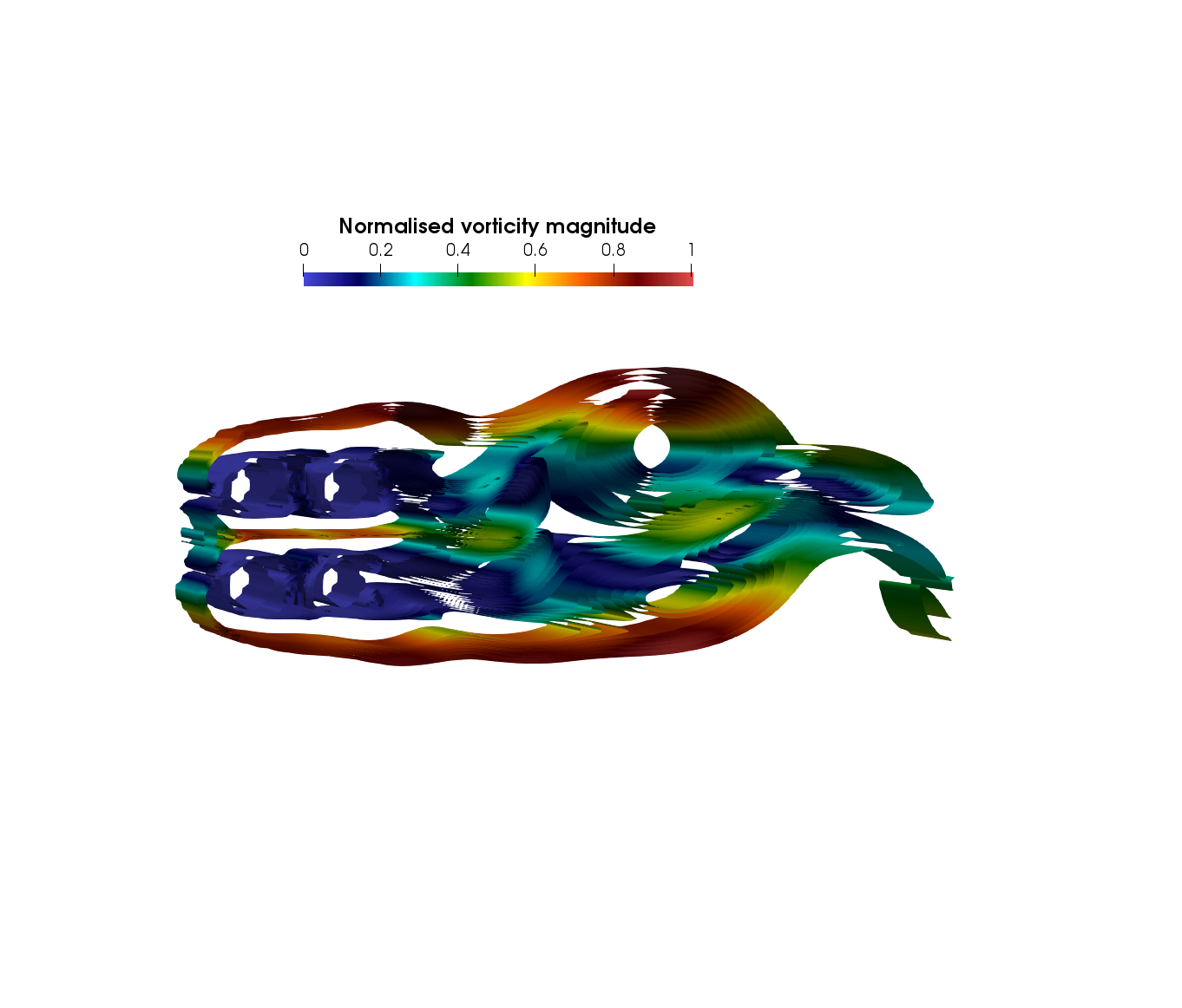}
           \caption{A three dimensional showing the time averaged vorticity magnitude as iso-surface for the case with spacing ratio $L/D=1.5$, were the difference in the wake behind the downstream cylinders is clear as discussed in the section before.}
          \label{fig:vort}
\end{figure}

Finally, Fig.~\ref{fig:vort} illustrates the vorticity iso-surface for the spacing ratio $L/D = 1.5$, highlighting the variation in the wake behind the downstream cylinders, as discussed in the preceding section.


\section{Summary}
In conclusion, this study has successfully developed and validated a novel computational approach for simulating three-dimensional viscous compressible turbulent flows at high Reynolds numbers. The integration of the Discretization Corrected Particle Strength Exchange (DC-PSE) method with a discrete subgrid-scale (SGS) filter offer a robust platform which is able to address the complexities associated with turbulence modeling, especially in intricate geometries.

By employing a Taylor expansion to define differential operators that closely approximate the convolution filter, and leveraging the Brinkman penalization technique for imposing boundary conditions implicitly, this research circumvents the need for altering the numerical method or computational domain. Furthermore, solving the Entropically Damped Artificial Compressibility (EDAC) formulation enables explicit simulations of the incompressible Navier-Stokes equations, enhancing the method's applicability to a wide range of high Reynolds number flow scenarios.

The accuracy of the method along side the convergence rate were demonstrated against several benchmark problems. The results showcased the method's ability to capture complex flow structures characteristic of turbulent flows.

While this study introduces significant advancements in the simulation of turbulent flows, it acknowledges certain limitations, particularly concerning the discrete Gaussian filter's order, the value of ($\epsilon$), and the effect of resolution on simulation accuracy and computational efficiency. Our findings indicate that the choice of filter order and the specific value of $\epsilon$ critically influence the quality of turbulence modeling, with higher-order filters and appropriately chosen $\epsilon$ values producing more accurate representations of energy dissipation and turbulent structures. Additionally, the resolution's impact on the simulations underscores the need for a careful balance between computational resources and the desired fidelity of the results.

\section{Acknowledgements} 
This work is funded by the Luxembourg National Research Fund (FNR) with the Core Junior grant lead by AO, ``A Numerical homogenisation framework for characterising transport properties in stochastic porous media'' (PorSol  C20/MS/14610324).

\bibliography{Barticle-3,Bbook,Brep,Bconf }

\end{document}